# Data Driven Modeling of Interfacial Traction Separation Relations using a Thermodynamically Consistent Neural Network

Congjie Wei[1], Jiaxin Zhang[2], Kenneth M. Liechti[3], Chenglin Wu[1*],

[1]Department of Civil, Architectural, and Environmental Engineering

Missouri University of Science and Technology, Rolla, MO, USA

[2]Computer Science and Mathematics Division

Oak Ridge National Laboratory, Oak Ridge, TN 37830

[3]Center for the Mechanics of Solids, Structures and Materials

Department of Aerospace Engineering and Engineering Mechanics

The University of Texas at Austin, Austin, TX, USA

* Corresponding author E-mail: zhangj@ornl.gov, wuch@mst.edu

## Abstract

For multilayer structures, interfacial failure is one of the most important elements related to device reliability. For cohesive zone modelling, traction-separation relations represent the adhesive interactions across interfaces. However, existing theoretical models do not currently capture traction-separation relations that have been extracted using direct methods, particularly under mixed-mode conditions. Given the complexity of the problem, models derived from the neural network approach are attractive. Although they can be trained to model data along the loading paths taken in a particular set of mixed-mode fracture experiments, they may fail to obey physical laws for paths not covered by the training data sets. In this paper, a thermodynamically consistent neural network (TCNN) approach is established to model the constitutive behavior of interfaces when faced with sparse training data sets. Accordingly, four conditions are examined and implemented here: (i) thermodynamic consistency, (ii) maximum energy dissipation path control, (iii) J-integral conservation, and (iv) mode-mix dependent toughness. These conditions are treated as constraints and are implemented as such in the loss function. The feasibility of this approach is demonstrated by comparing the modeling results with a range of physical constraints and several different input data sets. Moreover, a Bayesian optimization algorithm is then adopted to optimize the weight factors associated with each of the constraints in order to overcome convergence issues that can arise when multiple constraints are present. The resultant numerical implementation of the ideas presented here produced well-behaved, mixed-mode traction separation surfaces that maintained the fidelity of the experimental data that was provided as input. Some guidance is also provided on the desirable features of input data sets. The proposed approach heralds a new autonomous, point-to-point constitutive modeling concept for interface mechanics.

**Keywords:** Interface mechanics; Cohesive zone modelling; Traction-separation relations; Physics constrained neural networks; Machine learning; Bayesian optimization.

## Nomenclature

| Abbreviation | Meaning |
| --- | --- |
| MSE | Mean squared error |
| TC | Thermodynamically consistent |
| TC1 | Positive energy dissipation constraint |
| TC2 | Maximum energy dissipation constraint |
| TC3 | J-integral conservation constraint |
| TC4 | Mode-mix dependent toughness |
| TCNN | Thermodynamically consistent neural network |

| Symbol | Description |
| --- | --- |
| $\delta_n, \delta_t$ | Normal and tangential separation |
| $\sigma_n, \sigma_t$ | Normal and tangential traction |
| $\lvert\delta\rvert$, $\lvert\sigma\rvert$ | Separation, traction norms |
| $\phi$ | Phase angle |
| $J_n, J_t, J$ | Normal, tangential, and total J-integral |
| $d_n, d_t$ | Normal and tangential damage parameters |
| $\Gamma_n, \Gamma_t$ | Normal and tangential toughness |
| $\theta$ | Loading angle on the damage surface |
| $X, Y$ | Input and output data sets |
| $MSE_0$ | Loss function term corresponding to mean squared error |
| $MSE_1, MSE_2, MSE_3, MSE_4$ | Loss function terms incorporating the TC1, TC2, TC3 and TC4 constraints |
| $\lambda_0$ | Weight factor assigned to the mean squared error |
| $\lambda_1, \lambda_2, \lambda_3, \lambda_4$ | Weight factors assigned to the TC1, TC2, TC3 and TC4 constraints |
| $\epsilon_{phi}$ | Angle difference |
| $\mathrm{Vio}_1, \mathrm{Vio}_2, \mathrm{Vio}_3, \mathrm{Vio}_4$ | Violations of the TC1, TC2, TC3, TC4 conditions |
| $\overline{\mathrm{Vio}}_1, \overline{\mathrm{Vio}}_2, \overline{\mathrm{Vio}}_3, \overline{\mathrm{Vio}}_4$ | Normalized violations of the TC1, TC2, TC3, TC4 conditions |

# 1. Introduction

First proposed by Dugdale and Barrenblatt [1, 2], cohesive zone models have been widely adopted to describe nonlinear fracture processes, especially for interfacial fracture where fracture paths are essentially defined *a priori* [3-7]. Typically, cohesive zone models take pre-determined traction-separation relations as input either through cohesive elements or cohesive contact formulations in finite element codes to portray interfacial crack growth. Based on how traction-separation relations are implemented, cohesive zone models can be divided into those that are based on potentials or displacements [8]. The displacement approach typically uses specific forms of traction-separation relations including bilinear softening [9], cubic-polynomial [10], linear-softening [11] and exponential [12-14]. However, the physics behind the traction-separation relations that are used in this approach is not typically considered.

To ensure that the input traction-separation relations conform to the principles of thermodynamics, the potential-based models are based on the concept of cohesive energy potentials where the normal and tangential separations are adopted as independent variables. Cohesive tractions are defined as gradients of the cohesive energy potentials over separations for the softening stage. Needleman [15] pioneered the potential-based approach by introducing cubic-linear and exponential interfacial debonding potentials. Freed and Bank-Sills [16] extended the application to bi-material systems. Other types of potentials were also explored, including atomistic potentials relating metallic binding energies to lattice parameters [17], exponential-periodic potential adopting large tangential displacement jumps [7] and generalized exponential-periodic potential considering the relation between cleavage decohesion and dislocation nucleation [18]. These potential-based models are thermodynamically consistent and have a certain degree of flexibility to accommodate some of the behavior that has been observed in experiments. However, when dissipative interfacial failure mechanisms are involved, the potential-based approach often faces challenges in reconstructing the experimental data. Consequently, there is a need to develop a data-driven approach to model interfacial traction-separation relations that can capture the richness of the experimental data while remaining thermodynamically consistent. To this end, we seek a way to combine a data-driven approach, specifically deep learning, with thermodynamically consistent constraints in order to deliver a universal approach to model interfacial traction-separation relations that are extracted directly from experiments.

The deep learning approach has already been incorporated in computational fluid mechanics, where it is used to solve the governing partial differential equations of the fluid field [19-21] and constitutive modeling in plasticity [22-24]. The deep neural network generates potential solutions in terms of the displacement or velocity field that do satisfy the field equations. For determination of traction separation relations, models based on machine learning have also been proposed [25]. Wang et al. proposed a meta-modeling framework to generate constitutive model automatically using a Markov decision process [26]. However, these models have a strong dependence on analytically or numerically generated data rather than experimental results.

Efforts have also been made to take advantage of prior physical or mechanical knowledge in machine learning modeling processes [27]. By encoding the partial differential equations representing physical laws or governing equations into the machine learning algorithm, the training process was steered to converge towards the ground truth faster and more accurately [28, 29]. This approach could significantly reduce the amount of data needed for the training process and has proved to be a powerful tool in many fields [30-32]. However, most of the existing work has been focused on analytical or numerical models, where the amount and the type of data samples can be generated based on the training requirements. Furthermore, the governing equations are more directly incorporated and used in the construction of physics-informed machine learning models. This convenience might not be an option when it comes to situations where, for example, cost and time in obtaining the input data lead to sparse data sets. Another issue that can arise is the spatial-temporal randomness of the data. Moreover, if there are no analytical models available to assist with the accurate interpretation of the experimental data, as can be the case in modeling traction separation relations, then finding the best equations to incorporate in the machine learning model presents a challenge that is addressed in the current paper.

Building on these developments, the authors employ a deep neural network to learn from the traction-separation relations that are extracted from experiments along discrete mixed-mode loading paths, in order to generate traction surfaces in the mixed-mode space defined by the amplitude and phase of the interfacial separation vector. Thermodynamically consistent constraints that enforce positive energy dissipation, local maxima in energy dissipation, as well as the conservation of the J-integral and the often observed fact that interfacial toughness increases with mode-mix, are embedded in this deep neural network framework to form a thermodynamically consistent neural network (TCNN) for generating surfaces for traction-separation relations. The hyperparameter system, which is used for tuning the learning process, plays an essential role in the performance of neural networks. When inappropriate hyperparameter sets are adopted, convergence can be harder to achieve, and important features of traction-separation relations may be missed and lack robustness. Efforts have been made to develop algorithms for hyperparameter optimization, other than grid or random search concepts, which are time consuming and computationally costly optimization algorithms. As a result, Bayesian optimization [33, 34], tree structured Parzen estimators [35, 36], have been proposed to accelerate this process.

Here, we explore the effect of using different weight factors for the mean square error and thermodynamic constraints in order to generate normal and shear traction surfaces for a silicon/epoxy interface by employing mixed-mode fracture data generated by Wu et al. [37]. Metrics are established to demonstrate the effectiveness of Bayesian optimization in generating traction surfaces that retain the fidelity of the input data, while satisfying the selected thermodynamic consistency constraints.

The remainder of this paper is organized as follows. The theoretical underpinnings of our work are presented in Section 2. These are then implemented (Section 3) in the construction of a TCNN that incorporates four constraints and Bayesian optimization algorithms. The model verification is included in Section 4, where the model is verified based on the PPR model [38], validation data generated from a numerical analysis of a mixed-mode fracture problem, as well as the experimental data sets. The results are presented and discussed in Section 5 and conclusions can be found in Section 6. Detailed derivations, supplementary figures and tables are included as Supplementary Information at the end of the manuscript.

## 2. Traction-separation relations and thermodynamic consistency

### 2.1 Interfacial traction separation relations

For a two-layer structure undergoing interfacial fracture as shown in Figure 1a, tractions are provided by a cohesive layer between the two substrates. This cohesive layer does not have a thickness *per se* and is assumed to be homogeneous[1], which enables us to consider it as an assembly of identical springs connecting the two layers. The interfacial fracture process is then reproduced by the elongation and failure process of each spring along the interface. As shown in Figure 1a-b, for a stretched spring connecting the upper and lower layer, the normal and tangential tractions ($\sigma_n$, $\sigma_t$) change with the normal and tangential separations ($\delta_n$, $\delta_t$), which are defined by the change of the relative distances between the end points of the spring. Depending on the loading and boundary conditions, the ratio between the tangential and normal separations (i.e., the fracture mode-mix) varies. To quantitatively describe this relation, the vectorial separation is defined as the Euclidean norm, $|\delta|$, of the normal and tangential separation components. The mode-mix is represented by the phase angle, $\phi$, which is defined as the arctangent of the ratio between normal and tangential separations. Similarly, the vectorial traction is defined as $|\sigma|$. Thus

$$|\delta| = \sqrt{\delta_n^2 + \delta_t^2}, \tag{1a}$$

$$\phi = \operatorname{atan}\left(\frac{\delta_t}{\delta_n}\right), \tag{1b}$$

$$|\sigma| = \sqrt{\sigma_n^2 + \sigma_t^2}. \tag{1c}$$

[1] Note that this assumption is not required for the general framework that is being developed here. However it was the case in the data that will be used as input for the neural network.

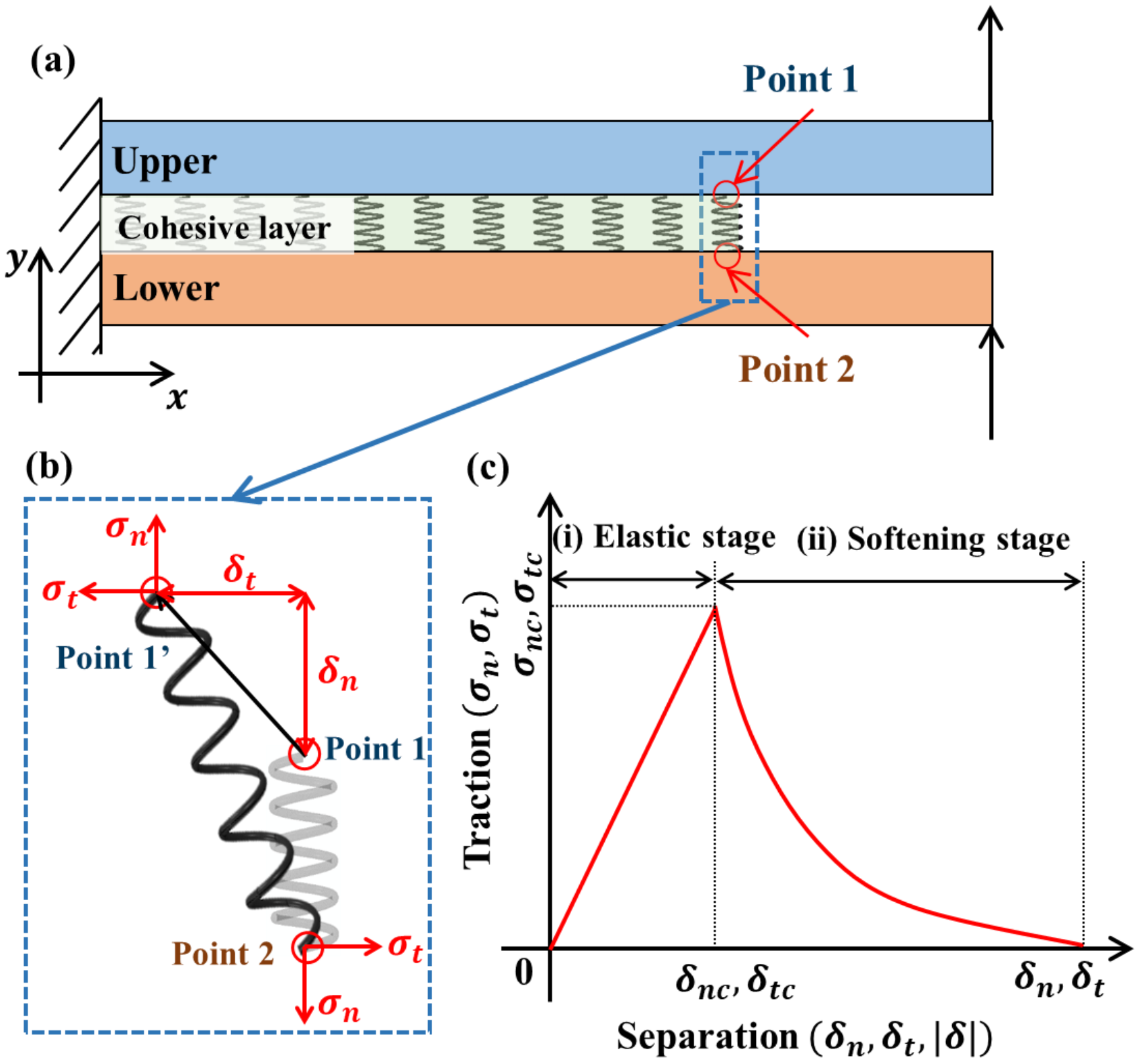

Figure 1: (a) Schematic of the specimen in an end loaded split experiment. (b) Tractions and separations for one spring. (c) A typical traction-separation relation with a two-stage response.

Traction-separation relations are the continuum representation of interfacial interactions. The normal and shear tractions, $\sigma_n(\delta_n, \delta_t)$ and $\sigma_t(\delta_n, \delta_t)$, respectively, on the interface are generally functions of the normal and tangential $(\delta_n, \delta_t)$ separations across the interface. Typically, a traction-separation relation has two stages: (i) the elastic stage during which the tractions increase along with the separations, and (ii) the softening stage where the tractions decrease as the separations continue to increase. An example of a traction-separation relation is shown in Figure 1c, where the elastic stage has a linear profile, and the softening stage follows an exponential one. To examine the features of traction-separation relation data in detail, we plot the mixed-mode traction-separation relations that were extracted for a silicon-epoxy interface [37] in Figure 2a-b.

Notice that the data for both normal and tangential relations are not smooth. Sharp peaks or discontinuities in gradients are commonly observed as shown in Figure 2c-d. These features make it unlikely that traction-separation relations can be represented by smooth functions, which are the hallmark of potential-based approaches [13, 38, 39].

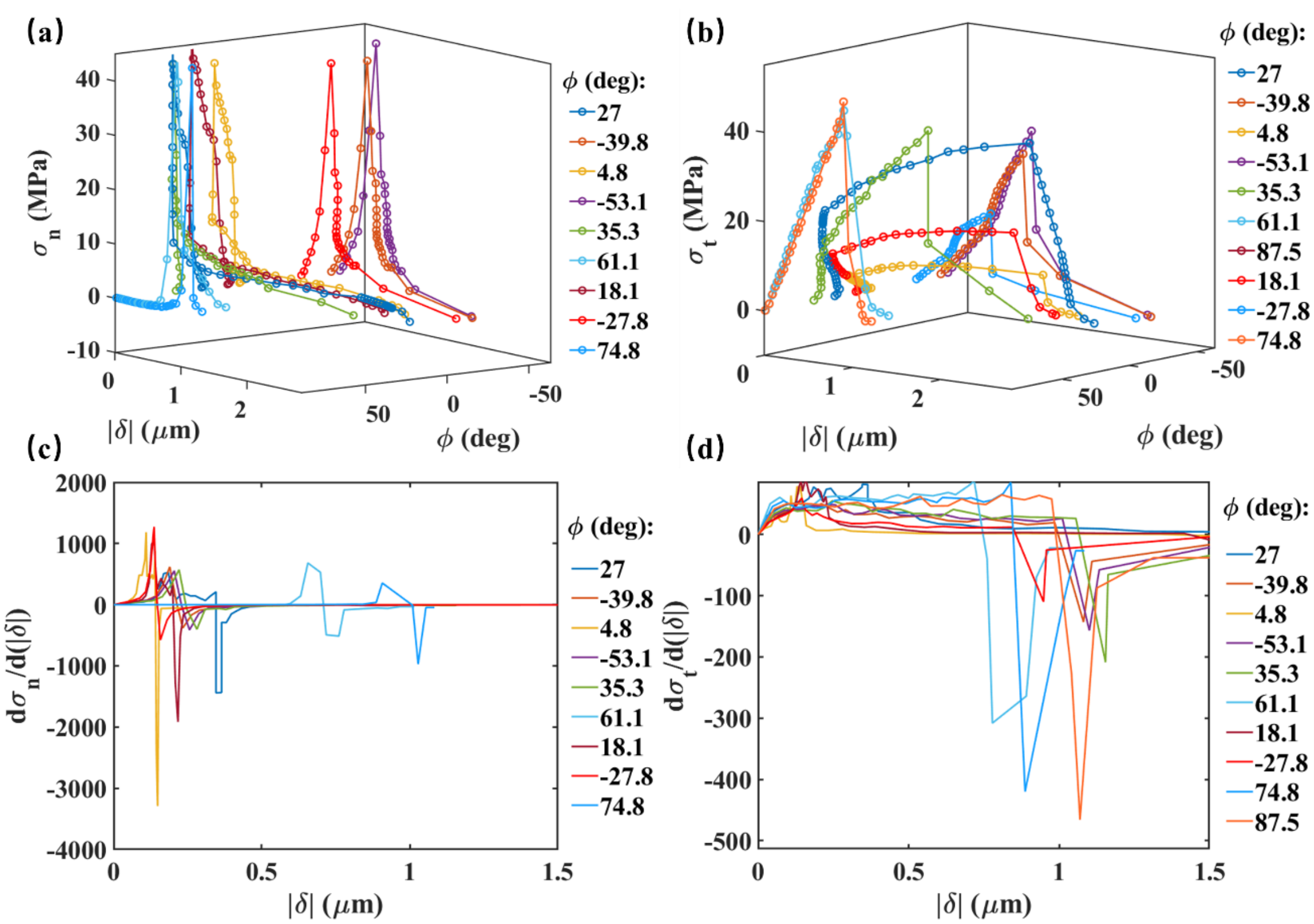

Figure 2. The (a) normal and (b) tangential components of the traction-separation relations for a silicon/epoxy interface in terms of total separation and phase angle. Gradient of (c) normal and (d) tangential tractions.

For instance, we explore the modeling of the experimental data with one of the most popular potential-based models: the Park, Paulino, and Roesler (PPR) model [38], which offers a large range of flexibility for modeling traction-separation relations using eight independent and adjustable parameters. The normalized modeling results are compared with the input experimental data in Figure 3. We note that the main inconsistency between the two data sets is mainly due to the smoothness of traction-separation relations obtained by the PPR model in contrast to the pointed profile of the measured traction-separation relations. The modeling processes, which make use of Monte Carlo algorithms, are provided in Section S2 of the Supplementary Information. The gradient surfaces and curves along fixed phase angle loading paths are provided in Figures S1-S2.

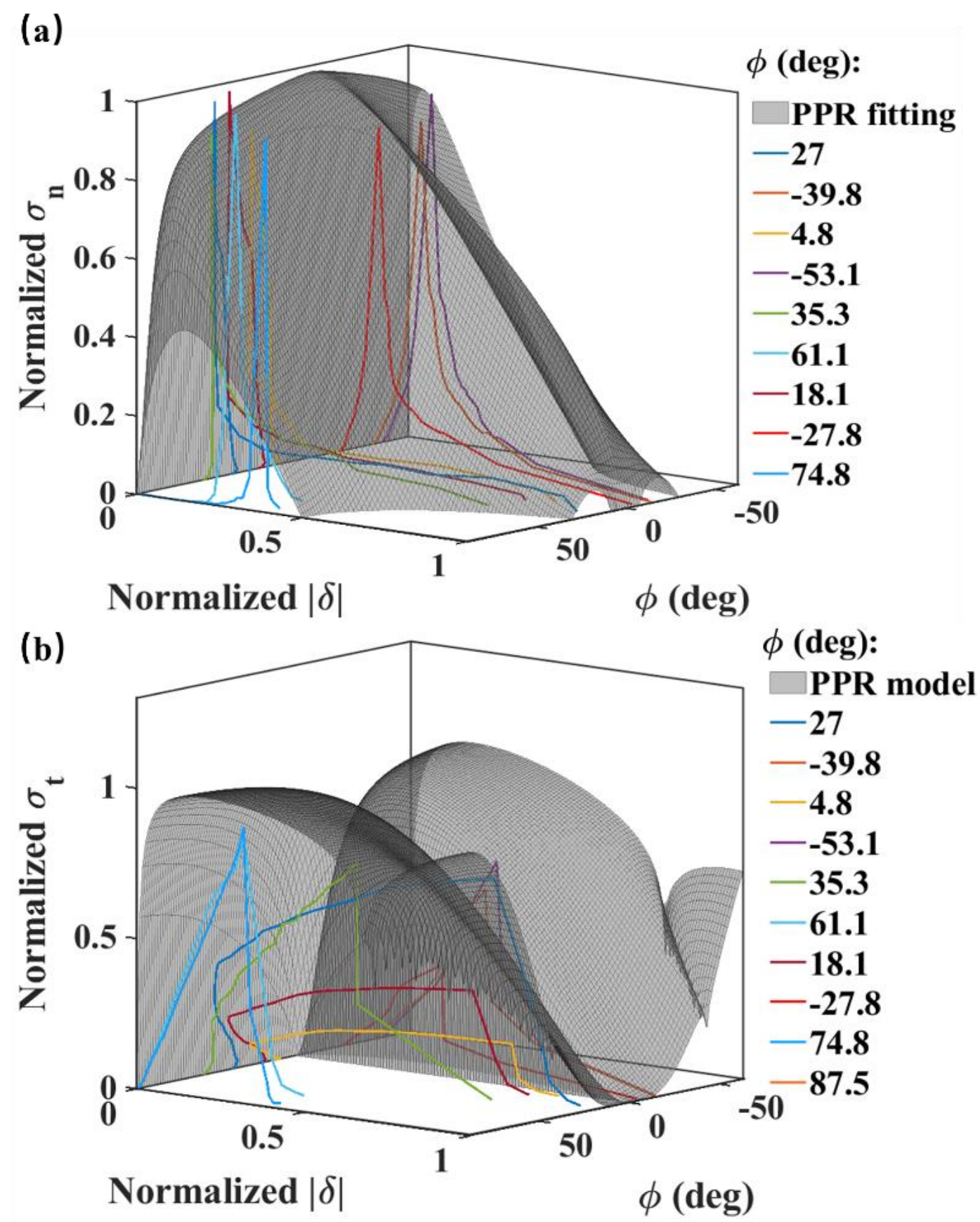

Figure 3. Normalized (a) normal and (b) tangential traction-separation surfaces that were modeled by the PPR model.

Certainly, there are other models [12] that might provide better fits to the experimental data. However, we argue that these models would suffer the same challenges as the PPR model. This is due to a fundamental conflict between these theoretically conceived modeling approaches and the experimental data. The smoothness required in the selection of functional forms to describe the interfacial energy is such that it cannot fully capture all the features of the experimental data. Therefore, there is a need for the development of a data-driven approach that allows traction-separation relations to be directly modeled from experimental data in order to preserve the richness in the data sets, while simultaneously capturing the underlying laws of physics. Here we develop such an approach by combining deep neural networks with conditions for thermodynamic consistency.

## 2.2 Thermodynamic consistency and toughness constraints

In this section we consider three conditions for thermodynamic consistency that the traction-separation surfaces obtained by deep neural networks should satisfy: (TC1), positive energy dissipation; (TC2), maximum energy dissipation; (TC3), conservation of the J-integral. The dependence of interfacial toughness on the fracture mode-mix is added as a fourth constraint (TC4). Each of the four conditions is described, quantitatively defined and prioritized based on certain features of the input data as well as other observations that are made from the experiments.

### 2.2.1 *TC1: Positive energy dissipation*

Fracture at interfaces is viewed as a process where the integrity of the interface is progressively lost due to the generation and propagation of defects, eventually yielding a fully cracked interface, as shown in Figure 4a. Both the epoxy and silicon remain elastic throughout the interfacial fracture experiments. Therefore, the energy dissipation is introduced as damage accumulates, which represented by the damage parameter, $d$, that ranges from 1 to 0 [40, 41], where $d = 1$ represents an intact interface and $d = 0$ for complete damage. Under monotonic loading, the energy dissipation associated with the accumulation of damage at the interface is an irreversible process, i.e., positive energy dissipation. This corresponds to the second law of thermodynamics. Based on damage mechanics concepts and the specific form that is adopted for cohesive zone modelling of interfacial fracture, the energy dissipation should always be positive, as described in the PPR model [14]. This also implies that the corresponding damage parameter (see Equation 2 for definition) should monotonically decrease during the loading process. This thermodynamic consistency requirement can be therefore formulated as follows.

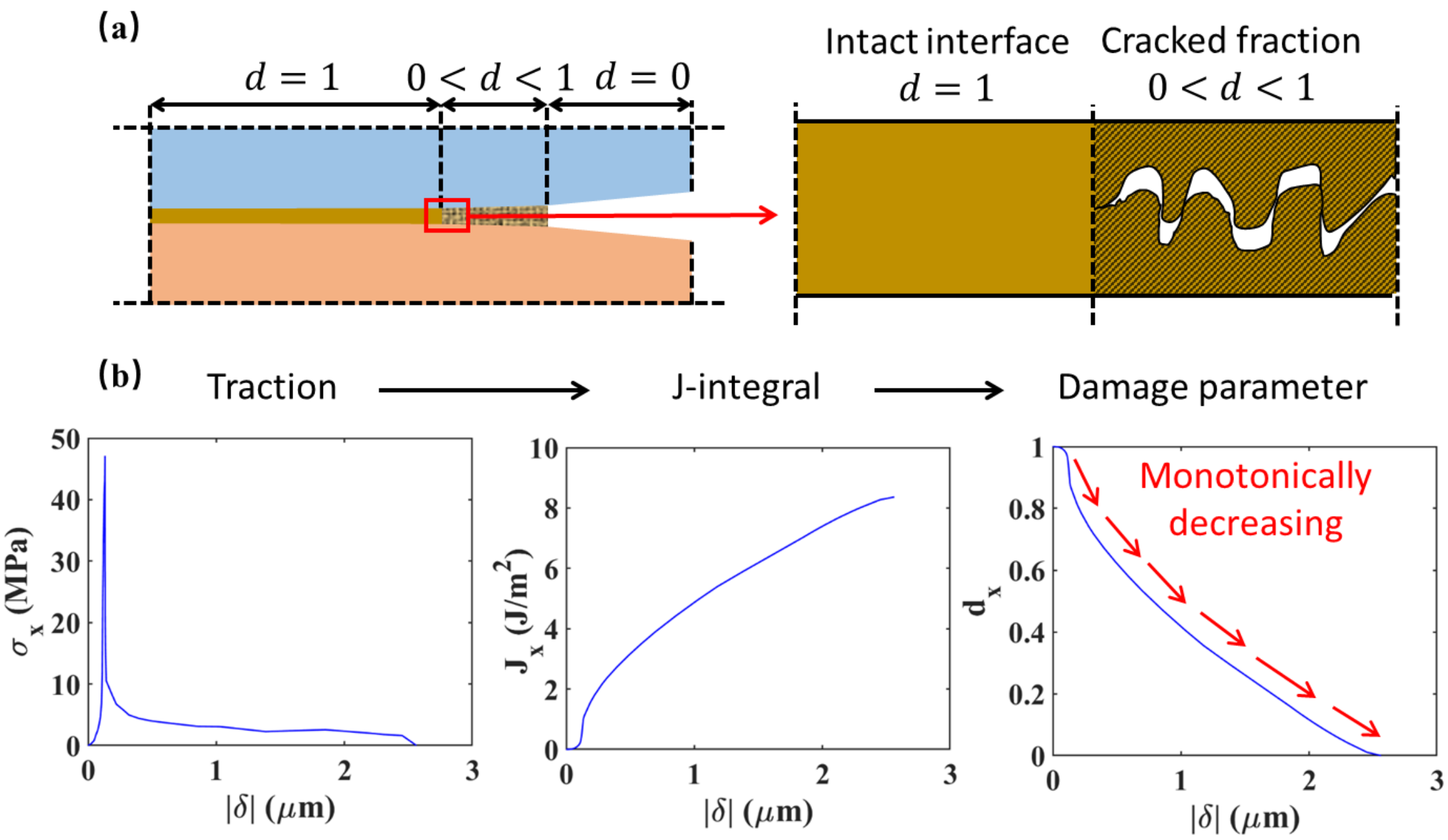

Figure 4. (a) Damage parameter of an interface. (b) Traction ($\sigma_i$, where ($x = n, t$) for normal or tangential) versus separation norm ($|\delta|$), J-integral ($J_i$) versus $|\delta|$ and damage parameter ($d_i$) versus $|\delta|$.

First, for an interface with pure mode I and mode II toughness values, $\Gamma_I$ and $\Gamma_{II}$, respectively, the damage parameters for the normal and tangential directions ($d_n$, $d_t$) are defined based on the J-integrals and toughness,

$$d_n(\delta_n) = 1 - \frac{J_n(\delta_n)}{\Gamma_I}, \tag{2a}$$

$$d_t(\delta_t) = 1 - \frac{J_t(\delta_t)}{\Gamma_{II}}, \tag{2b}$$

where $J_n$ and $J_t$ are the normal and tangential components of the J-integral that are obtained by integrating tractions with respect to separation:

$$J_n(\delta_n) = \int_0^{\delta_n} \sigma_n(\delta_n) d\delta_n, \tag{3a}$$

$$J_t(\delta_t) = \int_0^{\delta_t} \sigma_t(\delta_t) d\delta_t. \tag{3b}$$

It should be noted here that the input data for the approach being proposed here came from mixed-mode fracture specimens that satisfied a balance condition [37] that allows the J-integral to be decoupled into its normal and tangential components, in spite of the fact that the traction-separation relations themselves may be coupled.

The positive energy dissipation condition can be fulfilled by enforcing a monotonically decreasing damage parameter along with positive dissipation (Clausius-Duhem inequality) [42] as,

$$-\frac{\partial d_n}{\partial \delta_n} = \frac{\partial}{\partial \delta_n}\left(\frac{\int_0^{x=\delta_n} \sigma_n(x)dx}{\Gamma_n}\right) > 0, \tag{4a}$$

$$-\frac{\partial d_t}{\partial \delta_t} = \frac{\partial}{\partial \delta_t}\left(\frac{\int_0^{x=\delta_t} \sigma_t(x)dx}{\Gamma_t}\right) > 0, \tag{4b}$$

where $\Gamma_I = \max(J_n)$ and $\Gamma_{II} = \max(J_t)$.

As an example, the normal components of the traction, J-integral, and damage parameter for an experimentally obtained traction-separation relation at a nominal phase angle of 27° are shown in Figure 4b. The calculated J-integral increases while the corresponding damage parameter decreases monotonically, indicating that this experimentally obtained traction-separation relation satisfies the monotonically positive energy dissipation condition.

### 2.2.2 ***TC2: Local steepest descent***

When any friction or heat generated during interfacial fracture is negligible, the loading direction that results in the maximum separation will allow the dissipation rate to reach a local maximum when the interfacial separation reaches a local maximum. This is the case for the current work due to two reasons: (1) the strength of the interfacial interaction is lower than the yield strength of the epoxy [37, 43-45] so that no plastic dissipation is excited in the epoxy; (2) Furthermore, no significant polymer residual remained on the substrate after fracture, suggesting that the smooth fracture surface will minimize any frictional losses For the damage parameter surface shown in Figure 5a, loading along fixed phase angle paths results in the largest change in separation, making it the direction along which energy dissipation is fastest. For the point $d_{n2}^{(i)}$, loading along a fixed phase angle path and moving to point $d_{n2}^{(i+1)}$, results in the largest decrease of the damage parameter compared with the neighboring points (e.g., $d_{n1}^{(i)}$ and $d_{n3}^{(i)}$), where the phase angle does change. Here we define the loading angle as $\theta = tan^{-1}\left(\frac{d\phi}{d|\delta|}\right)$, as shown in Figure 5b. The corresponding gradient along all directions is calculated and presented in Figure 5c. It shows that the gradient reaches its minimum when $\theta = 0$, which is also the direction of the maximum interfacial separation.

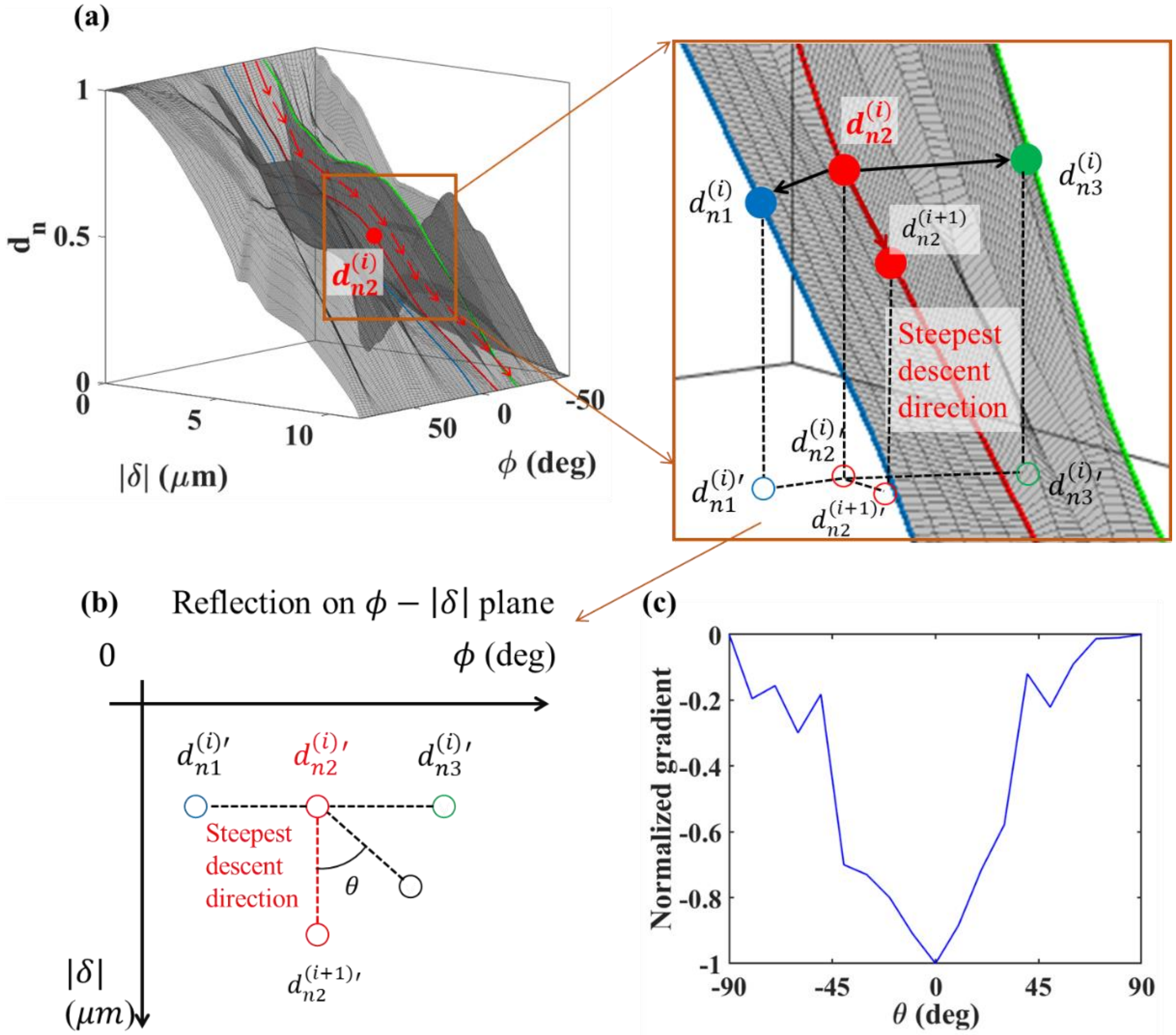

Figure 5. (a) Illustration of the local steepest descent condition. (b) definition of $\theta$ and (c) values of normalized gradient for any point (point $d_{n2}^{(i)}$ as illustration) along all directions.

This local steepest descent condition (TC2 condition) is applied to damage associated with both the normal and tangential components of the traction-separation relations and can be described as follows,

$$\left(\frac{\partial d}{\partial \delta}\right)_{\theta=0} \leq \frac{\partial d}{\partial \delta} \text{ at any } \theta \tag{5}$$

The proof of Equation 5 can be obtained using kinetic bond theories developed by Kramers and Bell [46-49] and recent applications to the silicon-epoxy interface [50, 51]. When the plastic work and frictional heat are ignored, we can assume the interfacial potential includes two components: (1) the elastic part coming from the external loading ($U_f = -f(t)\delta$), and (2) the interfacial potential ($U_b$). The magnitude of damage gradient is then a function of the gradient of potential with respect to the separation, i.e., $\left|\frac{\partial d}{\partial \delta}\right| = \frac{1}{\Gamma}\left|\frac{\partial U_b}{\partial \delta}\right|$. Under monotonic loading, we can obtain the following relationship between the separation velocity and damage gradient.

$$\left|\dot{\delta}\right| \propto \left|\dot{f}\right| = k_{off}\exp\left(-k_{off}t\right)\frac{KT}{\Gamma k_f}\left|\frac{\partial d}{\partial \delta}\right|, \tag{6}$$

where $K$ is Boltzmann's constant, T is temperature in Kelvin, $k_{off}$ is the bond off rate that can be approximately treated as a system constant, $t$ is time. Since the separation rate is proportional to the traction rate, the highest separation velocity at the time $t$ will result in the largest damage gradient. The detailed derivation of Equation 6 can be found in Section S2 of Supplementary Information.

#### 2.2.3 *TC3: Conserving the J-integral*

For interfacial failure processes, the total energy dissipation should equal the sum of the energy dissipation components along the normal and tangential separation directions if there is no other non-mechanical energy dissipation in the system (e.g., no heat generated from friction). For specimens that satisfy the balance condition and mixed-mode traction-separation relations along fixed phase angle loading paths, the total J-integral ($J$) can be decomposed into its normal and tangential J-integral components [50], as shown in Figure 6,

$$J(\sigma_n, \sigma_t, \delta_n, \delta_t) = J_n(\sigma_n, \delta_n) + J_t(\sigma_t, \delta_t) \tag{7}$$

Under this circumstance, the normal and tangential separations increase proportionally with a ratio that is the tangent of the phase angle,

$$\frac{\delta_t}{\delta_n} = tan\,\phi \tag{8}$$

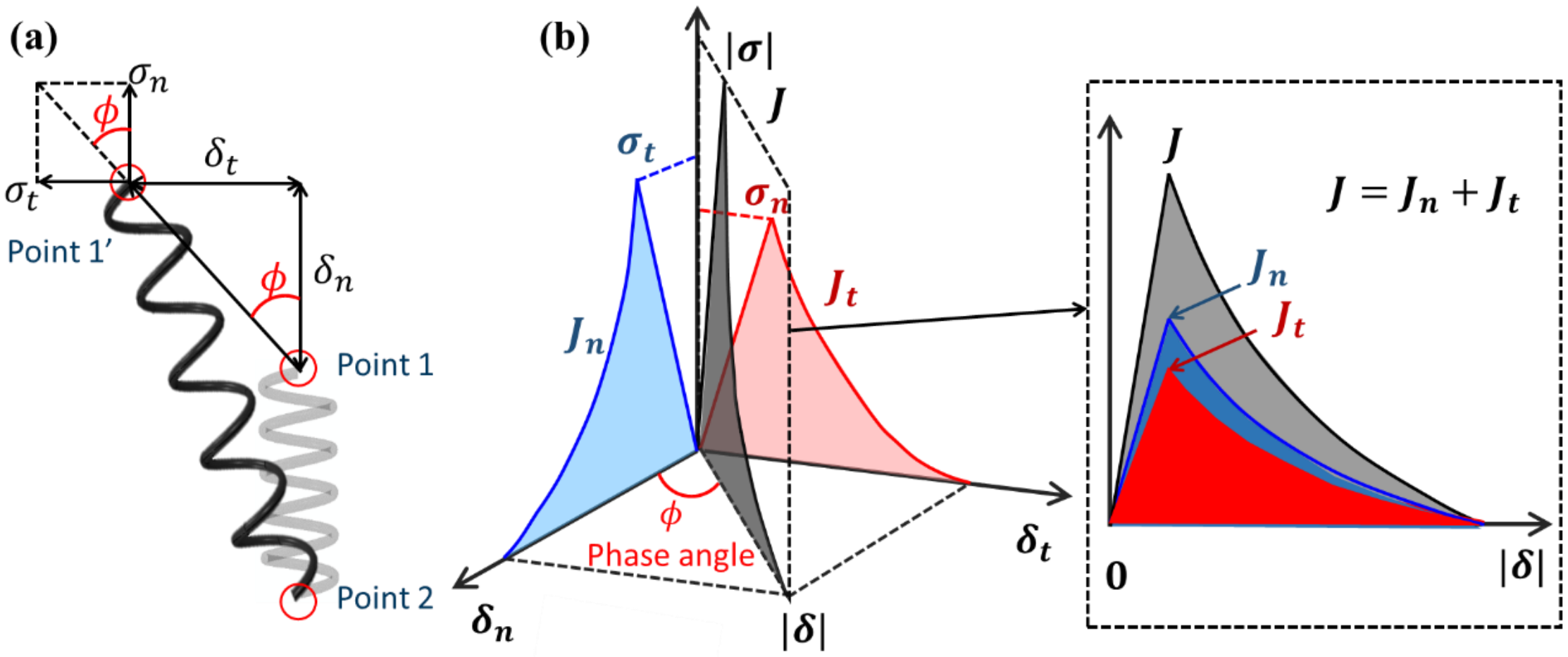

Figure 6. (a) Normal and tangential tractions and separations. (b) Conservation of the J-integral.

By replacing the normal and tangential components of the J-integral in Equation 7 with integrals of tractions with respect to separations, the relation between normal and tangential tractions is obtained and expressed as follows:

$$\frac{\sigma_t}{\sigma_n} = \frac{\delta_t}{\delta_n} = tan\,\phi\,. \tag{9}$$

This is the conservative loading path condition (TC3 condition) and its derivation from Equation 7 can be found in Section S2 of Supplementary Information.

### 2.2.3 *TC4: Mode-mix dependent toughness*

An additional constraint based on the variation of toughness with mode-mix is also considered here. It relates directly to traction-separation relations as the area underneath the curve and essentially provides an additional component for training the neural network. It can also be determined by tracking the steady state behavior of a cracked body. There are many examples in the literature beginning with the work reported in [52-55], where the toughness increases with as the shear component of the traction at the interface increases. Not all interfaces follow this behavior, but the silicon/epoxy data set [37] being considered here does. However, no generality is lost here as any variation of toughness can be implemented. Furthermore, because the balance condition was satisfied for all the specimens that were considered, the variation of the normal and shear components of the toughness could also be tracked as a function of mode-mix. Interestingly, while the total toughness increases with increasing shear, the normal and shear components of the J-integral decrease and increase, respectively, as the phase angle increases in magnitude (Figure 7). It is hypothesized that implementing this condition (TC4) will be helpful in improving performance of the neural network, particularly in the case of sparse data sets, where it essentially provides additional insight.

In reviewing all four constraints, first two conditions (TC1 and TC2) are developed based on the irreversible nature of energy dissipation in mechanical processes as stated in the second law of thermodynamics. This states that the process of elastic energy dissipated as heat or surface energy is irreversible. The third condition (TC3) assumes that the delamination occurs at the interface in the absence of crack face contact and any resulting frictional energy dissipation. This is not always the case [56-58], making this condition weaker than its first two counterparts. The fourth condition (TC4) is based on existing knowledge of a specific interface and may also be a stronger condition than TC3.

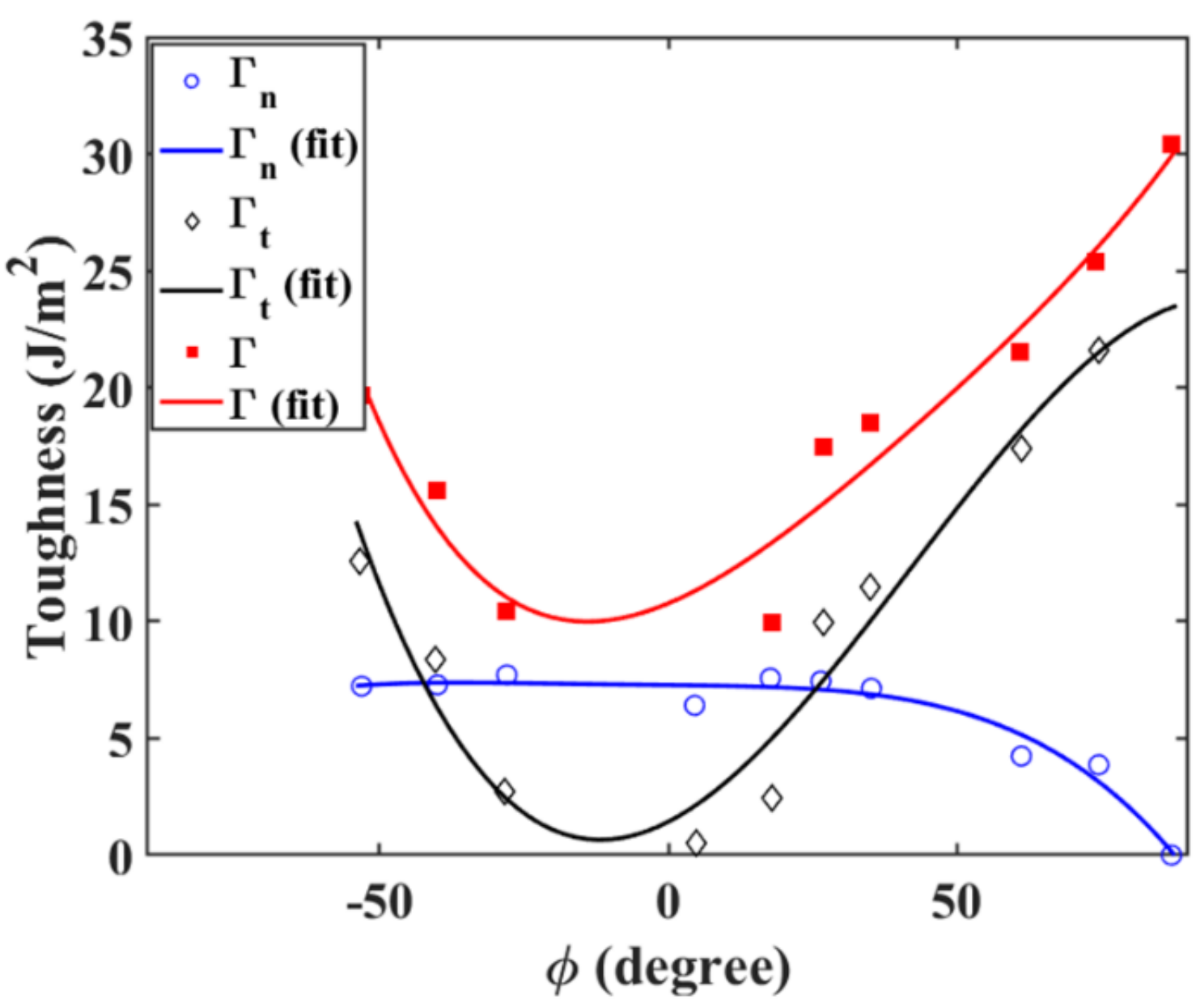

Figure 7. Interfacial toughness versus phase angle.

## 3. Thermodynamically consistent neural network

In this section, we choose to work with a deep neural network to formulate a data-driven approach to develop surfaces for traction-separation relations based on a limited number of data sets corresponding to the loading paths that were taken in a series of mixed-mode fracture experiments. To include thermodynamically consistent conditions in this model, each of the conditions (TC1, TC2, TC3, TC4) are embedded as loss function terms of the deep neural network to form the thermodynamically consistent neural network that is the basis of our approach. In addition, the importance of the experimental data and the thermodynamically consistent conditions are adjusted based on a Bayesian optimization algorithm, to balance any contradictions between experimental data and the thermodynamically consistent conditions.

### 3.1 Deep neural network construction

We first constructed the deep neural network with two hidden layers and 60 neurons for each layer as shown in Figure 8. The inputs are the separation norm ($|\delta|$) and phase angle ($\phi$) as defined in Equation 1a-b, whereas the outputs are the normal and tangential tractions ($\sigma_n, \sigma_t$) (Figure 8).

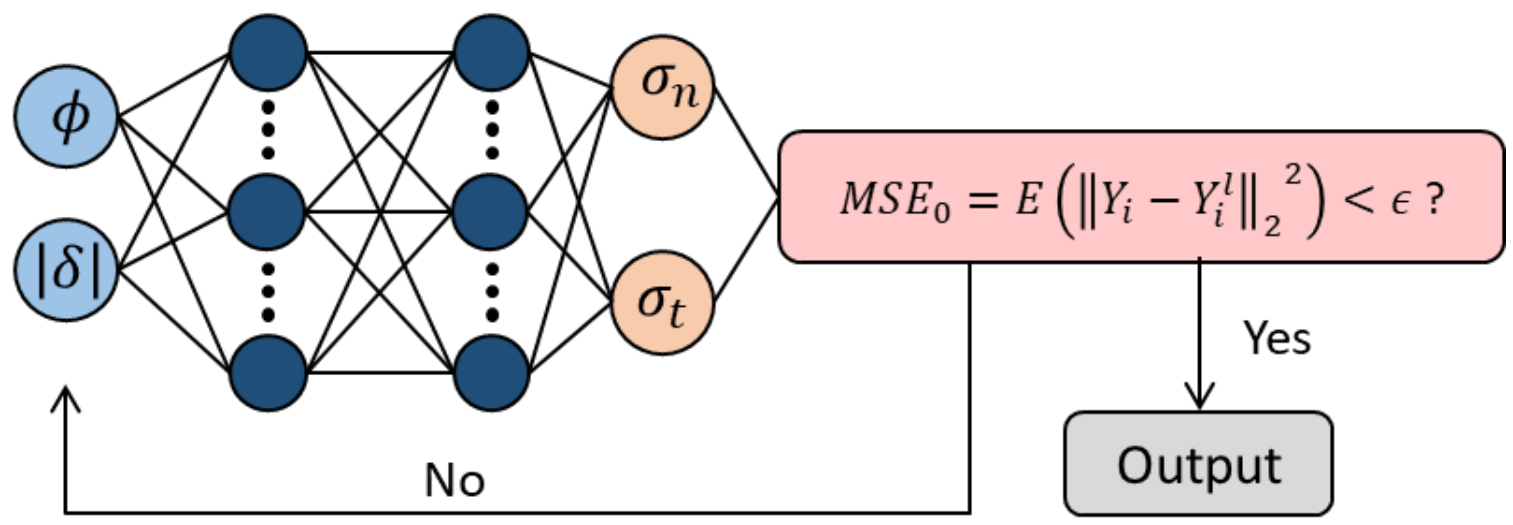

Figure 8. Deep neural network structure and optimization algorithm.

The input and output data are normalized by their maximum values to eliminate any bias introduced from relatively large differences in numerical values.

$$X_{norm}^{(i)} = \frac{X^{(i)}}{\max\limits_{i \in [1,n]} X^{(i)}}, \tag{10a}$$

$$Y_{norm}^{(i)} = \frac{Y^{(i)}}{\max\limits_{i \in [1,n]} Y^{(i)}}, \tag{10a}$$

where $X^{(i)}$and $Y^{(i)}$ are the $i^{\text{th}}$ values of the input and output data sets, $n$ is the number of elements in the dataset and $X_{norm}^{(i)}$ and $Y_{norm}^{(i)}$ are the normalized data for training the deep neural network. We use the hyperbolic tangent function as the nonlinear activation function and the Adam optimizer that has been widely used as an effective and fast optimization method [59]. The objective is to minimize the loss function that is embodied in the mean squared error and defined as:

$$MSE_0 = \mathrm{E}\left(\left\|Y_i - Y_i^l\right\|_2^2\right), \tag{11}$$

where $Y_i$ and $Y_i^l$ are $i^{\text{th}}$ points, respectively, of the true and projected output. The operator $\|\cdot\|$ represents the 2-dimensional Euclidean ($l_2$) norm. The training process is complete when 1,000,000 training epochs (defined as the number of iterations during the training process) are reached, or the loss function value is smaller than a specific threshold, specified here as $\epsilon = 10^{-3}$. It should be noted that the number of training epochs (1,000,000) is determined by experience. At this stage, the deep neural network that incorporates the thermodynamic consistency conditions outlined here should be in a position to develop surfaces for mixed-mode traction-separation relations across a wide range of mode-mixes from a discrete number of experiments along proportional paths in mode-mix space.

### 3.2 Implementation

Thermodynamically consistent conditions are added to the loss function as penalty terms in the deep neural network, in order to increase its effectiveness and accuracy. In this approach, the training process is one that searches for a solution that represents the training data sets sufficiently well. The loss function, which includes the mean square error, is used to determine how reasonable the prediction is and then incorporates this information in optimization algorithms for any subsequent training steps. By implementing thermodynamic consistency conditions as terms in the loss function, any trained models that violate them are penalized while the surviving models are, by definition, in compliance. This concept has been implemented in physics informed (or guided) neural networks and has worked well in a number of applications [27, 60, 61].

In this work, each of the four constraints are imposed on 10 loading paths along fixed phase angles ranging from $-60°$ to $90°$, shown as the red lines in Figure 9, taken at equal increments of $15°$. The interval along the $|\delta|$ coordinate is set at $0.1$ μm within a range of $[0, 3$ μm$]$. We refer to these paths as training paths, to differentiate them from the experimental paths. For each iteration along each training path, the loss function value is calculated automatically based on predicted traction-separation relation surfaces. This value is then used to adjust the TCNN and move on to the next iteration.

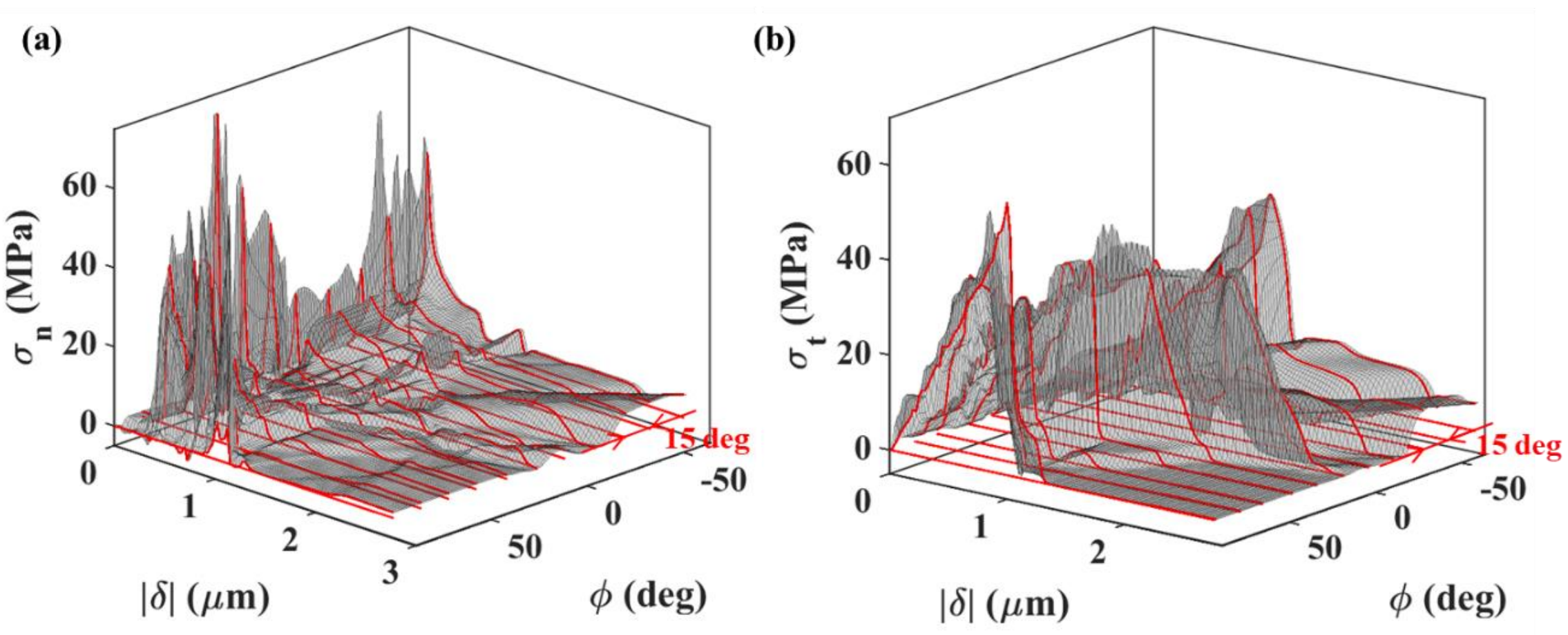

Figure 9. Illustration of fixed phase angle training paths on (a) normal and (b) tangential tractions.

### 3.2.1 ***Implementing the first thermodynamic constraint***

The first thermodynamic consistency condition (TC1) is imposed on training paths on the surfaces of both the output normal and tangential components of the traction-separation relations. For each training epoch, the gradient of the damage parameter for each training path is calculated. The Max function is then used to compare these values with 0 so that only the positive part, which is contrary to the TC1 condition, is considered

$$MSE_{1n}^{(j)} = max\left(\frac{\partial d_n}{\partial \delta_n}, 0\right) = \max_{i\in[1,n-1]}\left(\frac{d_n^{(i+1)}-d_n^{(i)}}{\delta_n^{(i+1)}-\delta_n^{(i)}}, 0\right), \tag{12a}$$

$$MSE_{1t}^{(j)} = max\left(\frac{\partial d_t}{\partial \delta_t}, 0\right) = \max_{i\in[1,n-1]}\left(\frac{d_t^{(i+1)}-d_t^{(i)}}{\delta_t^{(i+1)}-\delta_t^{(i)}}, 0\right), \tag{12b}$$

where $d_n^{(i)}$, $d_t^{(i)}$ are the $i^{th}$ normal and tangential damage parameters and $\delta_n^{(i)}$, $\delta_t^{(i)}$are separations on $j^{th}$ training path. The terms $MSE_{1n}^{(j)}$ and $MSE_{1t}^{(j)}$ are the normal and tangential components related to $j^{th}$ training path and $n$ is the total number of samples for each training path. Assuming that the normal and tangential damage parameters have the same weights, the new loss function ($MSE_1$) that incorporates the first constraint is formulated as,

$$MSE_1 = \frac{1}{2m}\sum_{j=1}^{m}\left(MSE_{1n}^{(j)} + MSE_{1t}^{(j)}\right), \tag{13}$$

where $m$ is the number of training paths. This loss function is averaged over the total number of points on the training paths.

### 3.2.2 *Implementing the second thermodynamic constraint*

The second thermodynamic consistency condition, TC2, (Equation 5) constrains the gradient vector direction along proportional phase angle loading paths. Numerically, this condition is reflected as a smaller gradient along the $|\delta|$ coordinate compared to the $\phi$ coordinate. Along each training path, the difference between the gradients along the $|\delta|$ and $\phi$ directions are calculated and compared with value 0 using Max functions via:

$$MSE_{2n}^{(i,j)} = \max_{\substack{i\in[1,n-1]\\ j\in[2,m-1]}} \left(\frac{\partial d_n^{(i,j)}}{\partial(|\delta|)} - \frac{\partial d_n^{(i,j)}}{\partial\phi}, 0\right) = \max_{\substack{i\in[1,n-1]\\ j\in[2,m-1]}} \left(\left(\frac{d_n^{(i,j)}-d_n^{(i+1,j)}}{\delta_n^{(i+1,j)}-\delta_n^{(i,j)}}\right) - \left(\frac{d_n^{(i,j)}-d_n^{(i,j+1)}}{\phi^{(i,j+1)}-\phi^{(i,j)}}\right), 0\right) + \max_{\substack{i\in[1,n-1]\\ j\in[2,m-1]}} \left(\left(\frac{d_n^{(i,j)}-d_n^{(i+1,j)}}{\delta_n^{(i+1,j)}-\delta_n^{(i,j)}}\right) - \left(\frac{d_n^{(i,j)}-d_n^{(i,j-1)}}{\phi^{(i,j)}-\phi^{(i,j-1)}}\right), 0\right) \tag{14a}$$

and

$$MSE_{2t}^{(i,j)} = \max_{\substack{i\in[1,n-1]\\ j\in[2,m-1]}} \left(\frac{\partial d_t^{(i,j)}}{\partial(|\delta|)} - \frac{\partial d_t^{(i,j)}}{\partial\phi}, 0\right) = \max_{\substack{i\in[1,n-1]\\ j\in[2,m-1]}} \left(\left(\frac{d_t^{(i,j)}-d_t^{(i+1,j)}}{\delta_t^{(i+1,j)}-\delta_t^{(i,j)}}\right) - \left(\frac{d_t^{(i,j)}-d_t^{(i,j+1)}}{\phi^{(i,j+1)}-\phi^{(i,j)}}\right), 0\right) + \max_{\substack{i\in[1,n-1]\\ j\in[2,m-1]}} \left(\left(\frac{d_t^{(i,j)}-d_t^{(i+1,j)}}{\delta_t^{(i+1,j)}-\delta_t^{(i,j)}}\right) - \left(\frac{d_t^{(i,j)}-d_t^{(i,j-1)}}{\phi^{(i,j)}-\phi^{(i,j-1)}}\right), 0\right), \tag{14b}$$

where $d_n^{(i,j)}$, $d_t^{(i,j)}$ and $\delta_n^{(i,j)}$, $\delta_t^{(i,j)}$ are the $i^{th}$ normal and tangential damage parameters as well as separations on $j^{th}$ training path. $MSE_{2n}^{(i,j)}$ and $MSE_{2t}^{(i,j)}$ are the normal and tangential components related to $i^{th}$ point on $j^{th}$ training path of this loss function term.

For data points obeying the second condition, this difference is negative, and the corresponding loss function term yields 0. For contradicting data points, the loss function term is positive and is minimized during the training process.

These constraints are imposed separately on normal and tangential traction-separation relation surfaces and play an equally important role in the total loss function term ($MSE_2$),

$$MSE_2 = \frac{1}{2mn}\left(\sum_{i=1}^{n-1}\sum_{j=2}^{m-1} MSE_{2n}^{(i,j)} + \sum_{i=1}^{n-1}\sum_{j=2}^{m-1} MSE_{2t}^{(i,j)}\right). \tag{15}$$

### 3.2.3 *Implementing the third thermodynamic constraint*

The third condition confines the ratio between the normal and tangential tractions to follow the training paths. The third loss function term ($MSE_3$) for each training path is defined as:

$$MSE_3^{(j)} = \frac{1}{n}\sum_{i=1}^{n} \left| \left( \frac{\sigma_t^{(i,j)}}{\sigma_n^{(i,j)}} - \tan\left(\phi^{(j)}\right) \right) \right|. \tag{16}$$

The terms $\sigma_n^{(i,j)}$, $\sigma_t^{(i,j)}$ are the $i^{\text{th}}$ normal and tangential tractions of $j^{\text{th}}$ training path. $MSE_3^{(j)}$ is the loss function related to the $j^{\text{th}}$ training path. This constraint differs from the first two in that it is an equality as opposed to an inequality. The assembled total loss function ($MSE_3$) term corresponding to the third constraint is written as:

$$MSE_3 = \frac{1}{m}\sum_{j=1}^{m} MSE_3^{(j)}. \tag{17}$$

### 3.2.4 *Implementing the toughness constraint*

The fourth condition embodies the monotonicity of toughness with mode-mix. The toughness at any phase angle is extracted from the predicted J-integrals with $|\delta| = |\delta_f|$, where the $|\delta_f|$ is a relatively large absolute separation where the J-integral values reach their steady state. In surveying all the experimental data, the value $|\delta_f|$ is set to be 2.8 μm for this particular interface, thereby ensuring that tractions of all experimental paths reduced to 0. The corresponding normal and tangential loss function terms for TC4 are then formulated as,

$$Loss_{4n,i} = \max_{i\in[1,n-1]} \left( \frac{J_n^{(i+1)} - J_n^{(i)}}{|\phi^{(i+1)}| - |\phi^{(i)}|}, 0 \right), \text{ for } |\delta| = |\delta_f| \tag{18a}$$

$$Loss_{4t,i} = \max_{i\in[1,n-1]} \left( \frac{J_t^{(i)} - J_t^{(i+1)}}{|\phi^{(i+1)}| - |\phi^{(i)}|}, 0 \right), \text{ for } |\delta| = |\delta_f| \tag{18b}$$

The total loss for this TC4 condition is then,

$$MSE_4 = \frac{1}{2m} \sum_{j=1}^{m} \left( MSE_{4n}^{(j)} + MSE_{4t}^{(j)} \right) \quad (19)$$

### 3.2.5 *Weighted loss functions for the thermodynamically consistent neural network*

The loss function terms, with the conditions TC1-TC4 embodied as constraints (Equations 13, 15, 17, 19 respectively), are implemented in the deep neural network so as to construct a thermodynamically consistent neural network. The schematic for this network is shown in Figure 10. The deep neural network uses the same architecture as was discussed in Section 3.1: it has two hidden layers with 60 neurons for each hidden layer.

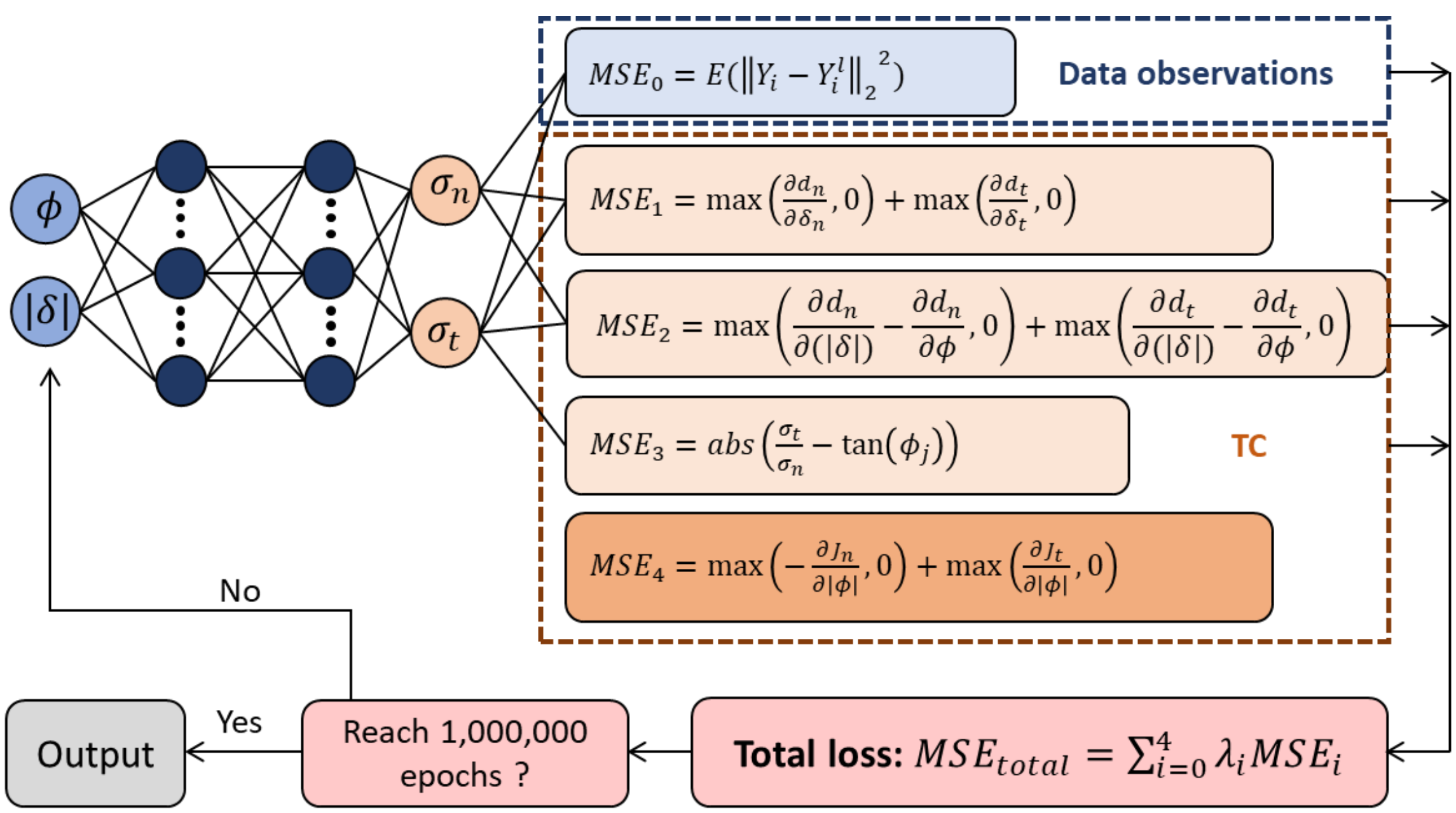

Figure 10. Schematic of a thermodynamically consistent neural network for traction surfaces

Different weight factors $\lambda_i$ were considered in order to investigate the effect of the imposed thermodynamically consistent constraints on the performance of the proposed neural network. Four different neural networks (designated TCNN1 to TCNN4) were compared by considering four sets of weight factors. The active constraints and weight factors for all four sets are listed in Table 1, bearing in mind that $\sum_{i=0}^{4} \lambda_i = 1$.

Table. 1 Weight factor setting

| | $\lambda_0$ | $\lambda_1$ | $\lambda_2$ | $\lambda_3$ | $\lambda_4$ | Active TC constraints |
|---|---|---|---|---|---|---|
| TCNN0 | 1.000 | 0.000 | 0.000 | 0.000 | 0.000 | N/A |
| TCNN1 | 0.900 | 0.100 | 0.000 | 0.000 | 0.000 | TC1 |
| TCNN2 | 0.800 | 0.100 | 0.100 | 0.000 | 0.000 | TC1, TC2 |
| TCNN3 | 0.780 | 0.100 | 0.100 | 0.0200 | 0.000 | TC1, TC2, TC3 |
| TCNN4 | 0.680 | 0.100 | 0.100 | 0.0200 | 0.100 | TC1, TC2, TC3, TC4 |

From TCNN1 to TCNN4, the weight factor for $MSE_0$ term ($\lambda_0$) is reduced and redistributed to the thermodynamic consistency conditions step by step. An accurate prediction of the traction-separation relations along the paths taken in the series of experiments was still given the highest priority. This meant that $\lambda_0$ was always greater than 0.5 in order to ensure that the main features of the data from the experiments were captured. The constraints TC1 and TC2 were treated with equal importance and thus were assigned the same weight factor. Being an equality, the third constraint is inherently stronger than the inequalities that are embodied in the first two constraints. If all four constraints are assigned the same weight factor, the third one would therefore dominate the process. Accordingly, $\lambda_3$ was assigned smaller values in order to balance the contribution from each of the three constraints. Due to the specificity of the fourth constraint and the fact that it is an inequality, $\lambda_4$ was assigned as the same weights as TC1 and TC2.

### 3.3 Bayesian optimization

In the previous section, the certain weight factors were selected in order to demonstrate their effect in a limited manner. This raises a question as to what the best rationale is for selecting the various weight factors. Here we selected Bayesian optimization [62, 63] to automatically adjust the weight factors, so that the relative importance of mean square error and thermodynamic consistency terms are optimized, and a minimum total loss function can be reached.

In many cases, experimental data is accompanied by systematic errors and noise, which can lead to localized conflict between experimental results and thermodynamic consistency conditions. When used as training data sets in the TCNN models, these factors can significantly affect the projected traction-separation surfaces and lead to portions of the surfaces that violate thermodynamic consistency. Moreover, this conflict can interfere with the training processes within the TCNN, which can result in issues such as excessive training for convergence or poor representations of the experimental data. By using automatically selected weight factor adjustment algorithms for optimization, a minimum loss function value can be reached in an optimal manner. As a result, the projection of traction-separation relations may adjust portions of the input data to satisfy thermodynamic consistency when conflicts are identified, while still capturing the main features of the input data.

The Bayesian optimization algorithm that performs global minimization of unknown functions with multiple degrees of freedom [62, 63], is adopted for the optimization. Using knowledge obtained from some prior sampling over the space of this unknown function, the Bayesian optimization algorithm estimates the areas with highest possibility to generate most promising results. In this way, instead of conducting random or grid searches where results should be obtained for every set of parameters, the Bayesian optimization algorithm provides more optimal searches and reduces the number of trial attempts by taking advantage of prior knowledge and the calculated likelihood of the unknown area. Our implementation is conducted within the confines of the Trust-region Bayesian optimization (TuRBO) algorithm [34], which solves for global optimization of large-scale high-dimensional problems using a local probabilistic approach.

In integrating the Bayesian optimization algorithm with the TCNN described earlier, the set of the weight factors is determined such that it minimizes the overall loss function value while the following actions are taken: (1) the weight factors for all terms in the loss function are relaxed and allowed to vary within a reasonable range; (2) the values of the weight factors used in TCNN4 were upper bounds for those selected in the optimization process.

The implementation of TCNN framework with Bayesian optimization is shown in Figure 11 with the following steps.

Step 1. The normalized training data and the randomly initiated weight factors are input for the first training step.

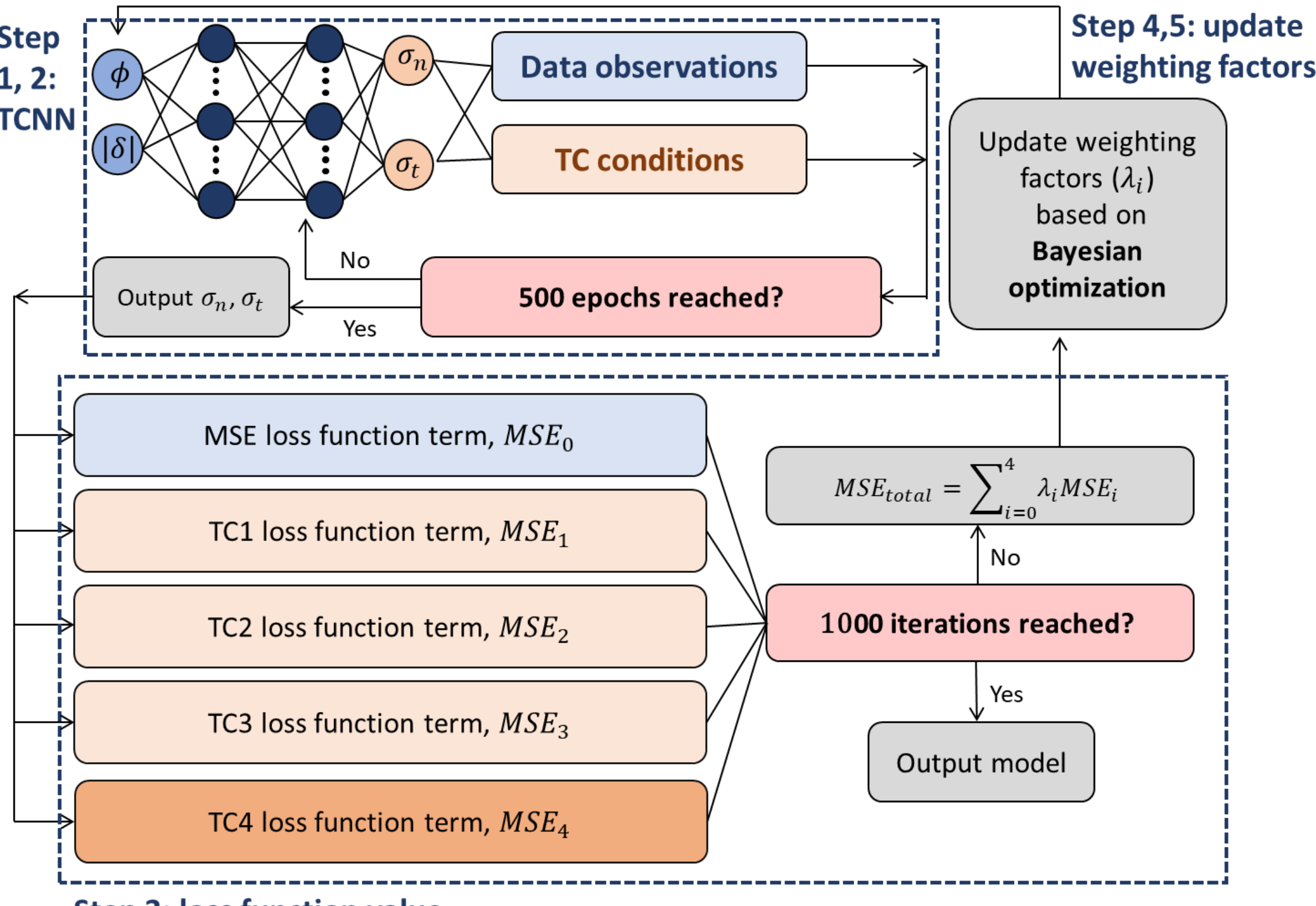

Figure 11. Bayesian optimization of weight factors.

Step 2. TCNN is trained for 500 epochs; this relatively low epoch number reduces the time required for each iteration of the Bayesian optimization.

Step 3. Upon the completion of the TCNN training process, values of loss function terms corresponding to MSE (based on Equation 11) and TC1-4 conditions (based on Equation 13, 15, 17 and 19) are calculated. The total loss function value is then obtained from Equation 20.

$$MSE_{total} = \sum_{i=0}^{4} \lambda_i MSE_i. \quad (20)$$

Step 4. The weight factors are updated based on Bayesian optimization criteria.

Step 5. The updated weight factors are input to the TCNN to start the next iteration. This cycle continues until the maximum budget (1000 iterations, determined by experience) is reached.

For each iteration in the Bayesian optimization algorithm, the weight factors are optimized under two constraints: (1) Each weight factor varies within a range defined by lower and upper bounds ($\lambda_i \in [\lambda_i^{lower}, \lambda_i^{upper}], i = 0, 1, 2, 3, \text{or } 4$). This allows the user to rank the constraints while avoiding weight factors that are smaller than zero. (2) The weight factors must satisfy the condition $\sum_{i=0}^{4} \lambda_i = 1$. The updated weight factor set is transferred to the TCNN for the next training step. The end of the optimization process is reached when the maximum number (e.g., 300, determined by experience) of iterations is reached.

## 4. TCNN model validation

The validation of the approach is first conducted using input data generated analytically using the PPR model (Park, Paulino et al. 2009). The actual experimental data that was employed in this study consisted of ten different mixed-mode loading paths with a total of 1233 data points. Two strategies are then explored for dividing this experimental data into validation and training data sets. Once the best strategy was established, we then optimized the hyperparameters for the neural network. It should be noted that for all cases presented in this section, the weight factors are fixed with the values shown in Table 1.

### 4.1 Validation with analytical input data

Input data for the TCNN model (TCNN3, weight factors can be found in Table 1) was first provided by the PPR model [38] because it was then a simple matter to analytically generate test cases for the TCNN model that were outside the input data set. The training data sets and the validation data sets were all distinct because separate strategies were used to generate each data set. The training data sets were generated for loading paths at with 10-degree intervals in phase angle while the validation data sets were generated via a Latin hypercube sampling approach along nine other loading paths and an overlapping one at 90°. The normal and shear traction surfaces that were used for training are shown in Figures 12 and 13, respectively. In each case that data is compared with the trained and validation data sets that were produced by the TCNN. The prediction error for the validation data sets were 1.39 % $\pm$ 0.38 % and 0.47 % $\pm$ 0.36 % for the normal and shear components, respectively. This demonstrates the capability of TCNN model to reconstruct traction surfaces that are generated with an analytical cohesive model.

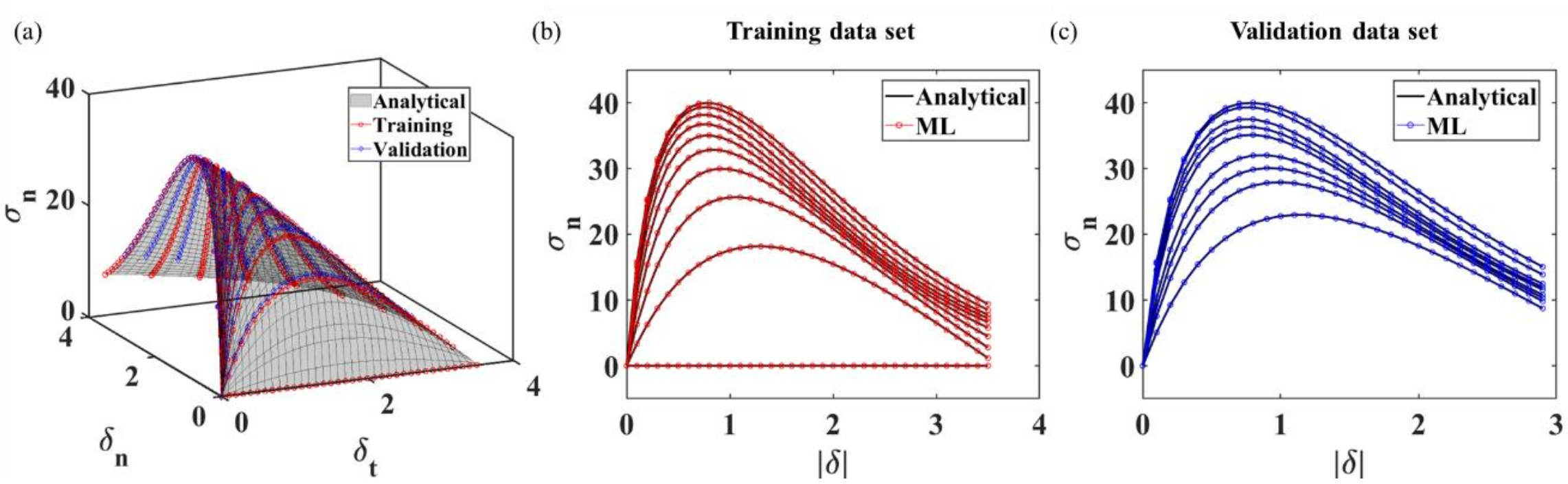

Figure 12. (a) The analytically generated normal traction surface from a PPR model. The red and blue symbols reflect the data that was used for training and validation, respectively; (b) Comparing trained and analytical data sets for $\phi = 0,10,20,30,40,50,60,70,80,90$; (c) Comparing validation and analytical data sets for $\phi = 10.7, 23.2, 30.9, 38.3, 48.4, 59.3, 64.1, 73.6, 84.5, 90$.

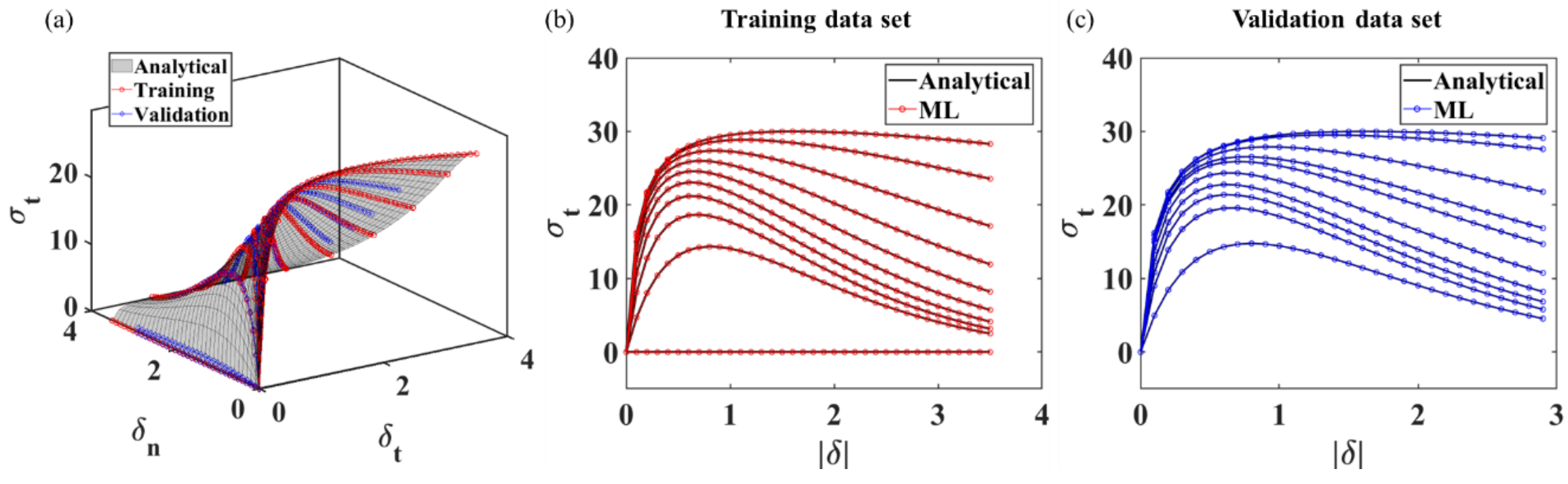

Figure 13. (a) The analytically generated shear traction surface from a PPR model. The red and blue symbols reflect the data that was used for training and validation, respectively; (b) Comparing trained and analytical data sets for $\phi = 0,10,20,30,40,50,60,70,80,90$; (c) Comparing validation and analytical data sets for $\phi = 10.7, 23.2, 30.9, 38.3, 48.4, 59.3, 64.1, 73.6, 84.5, 90$.

In addition to the basic validation, we also determined the sufficient data needed for accurate TCNN predication in terms of number of load path (or phase angles). For this purpose, we explored the validation using three series with six, seven, and eight loading paths included for each series, respectively, and based on the data produced analytically from the PPR model. Each loading path follows a fixed phase angle ($\phi$) with the absolute separation ($|\delta|$) ranging from 0 to 3.5 with a 0.1 interval. For each of the series, data along each path is selected as the validation data while the rest are used for TCNN training. Considering that $\sigma_n$ or $\sigma_t$ remains 0 along $\phi = 0 \; or \; 90$ paths, all paths excluding the zero traction cases are used as validation data sets. Modeling results for series with six paths are shown in Figures S3-8, while those for series with seven and eight paths are shown in Figures S9-15 and Figures S16-22 respectively. The TCNN validation results for the phase angle of 12.9° is shown in Figure 14 for eight loading paths with an error of 1.10 % for normal tractions and 8.22 % for tangential tractions. The overall prediction errors for all series are shown in Figure 15.

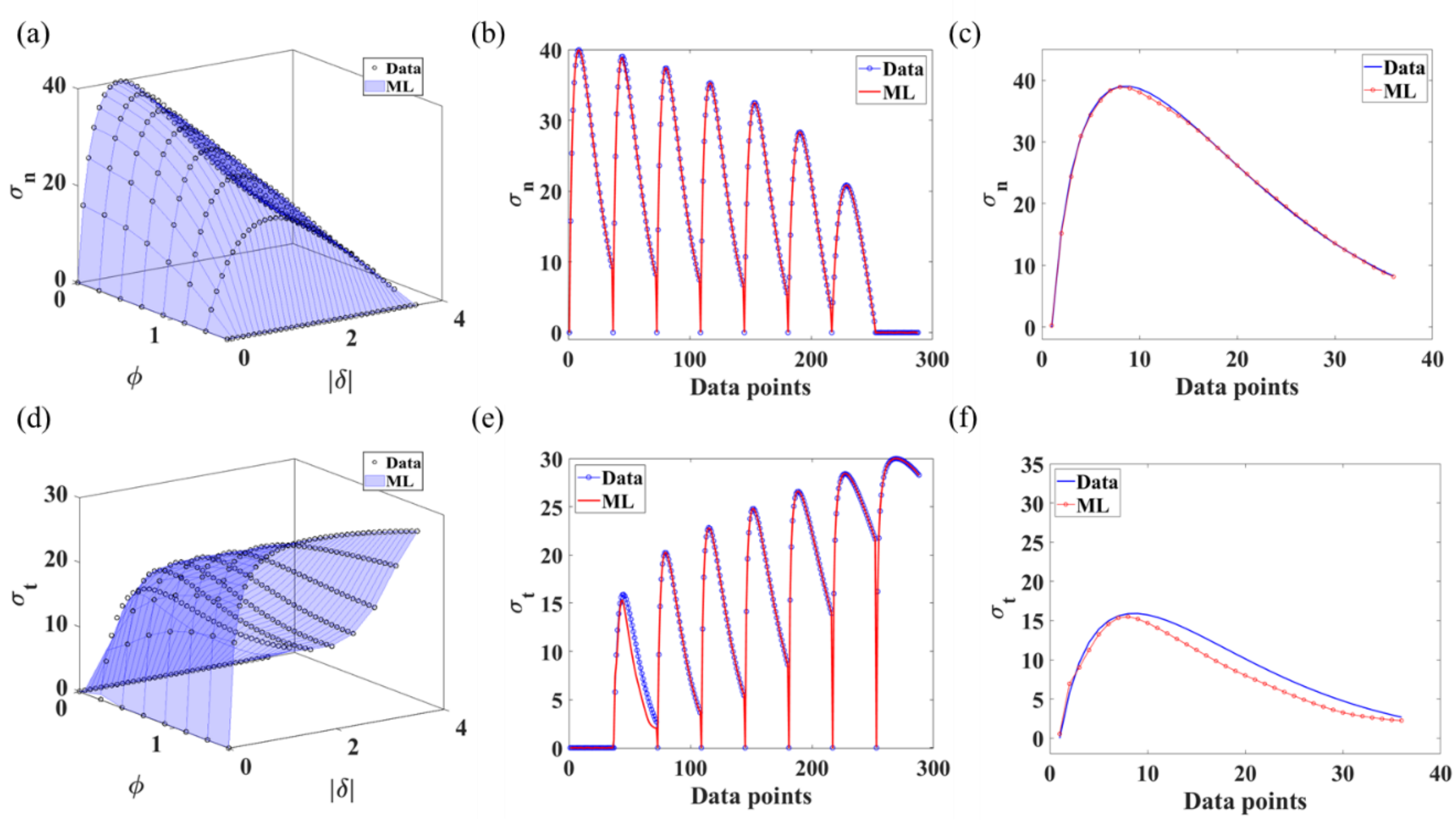

Figure 14. Sample modeling results for normal and tangential stresses obtained from the PPR analytical model with eight paths, where 2nd path is used as validation data. (a) Predicted normal

traction surface. Comparison between data and ML predicted results for normal stresses of (b) all data sets and (c) validation data sets. (d) Predicted tangential traction surface. Comparison between data and ML predicted results for tangential tractions of (e) all data sets and (f) the validation data set.

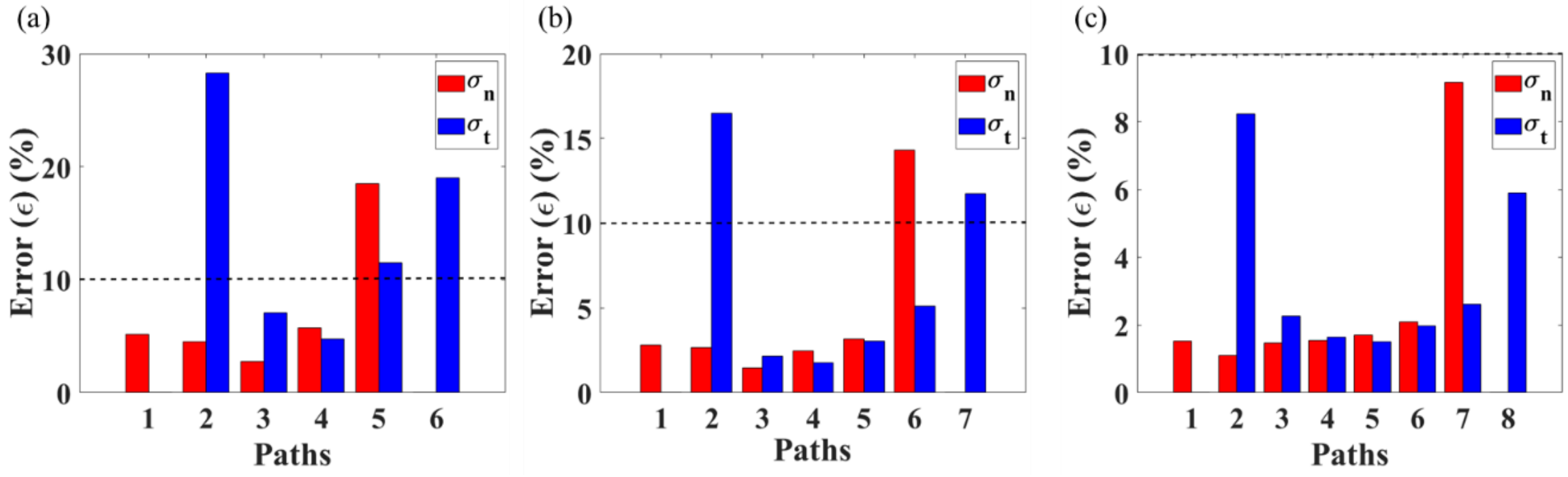

Figure 15 Prediction error for (a)-(c) 6-8 paths series.

From the results, we observed that the predication errors gradually reduce, and drop below 10% when the number of loading paths becomes 8. Therefore, we expect that a minimum of eight loading paths with fixed phase angles will be needed to achieve an accurate prediction.

### 4.2 Validation with data generated from numerical modeling

In this section, the TCNN model (TCNN3) is validated based on data generated from finite element analyses (ABAQUS) of an end loaded split specimen (Fig.16a), where the beams are assumed to be elastic and the interlayer is modeled with cohesive elements. Bilinear, normal and shear traction-separation relations are adopted for the interlayer (Fig. 16b). The selected parameters for the cohesive elements are shown in Table S1. The applied normal and tangential displacements $(U_x, U_y)$ of the far-end are selected to create cases with a range of phase angles. Eleven loading paths are considered where the phase angles are $0.00, 14.30, 24.58, 35.71, 51.17, 56.24, 64.55, 72.13, 79.24, 85.22, 90.00$ degrees. Numerical

modeling details regarding the initiation and damage growth criteria is provided in Section S3 and Table S1-S2.

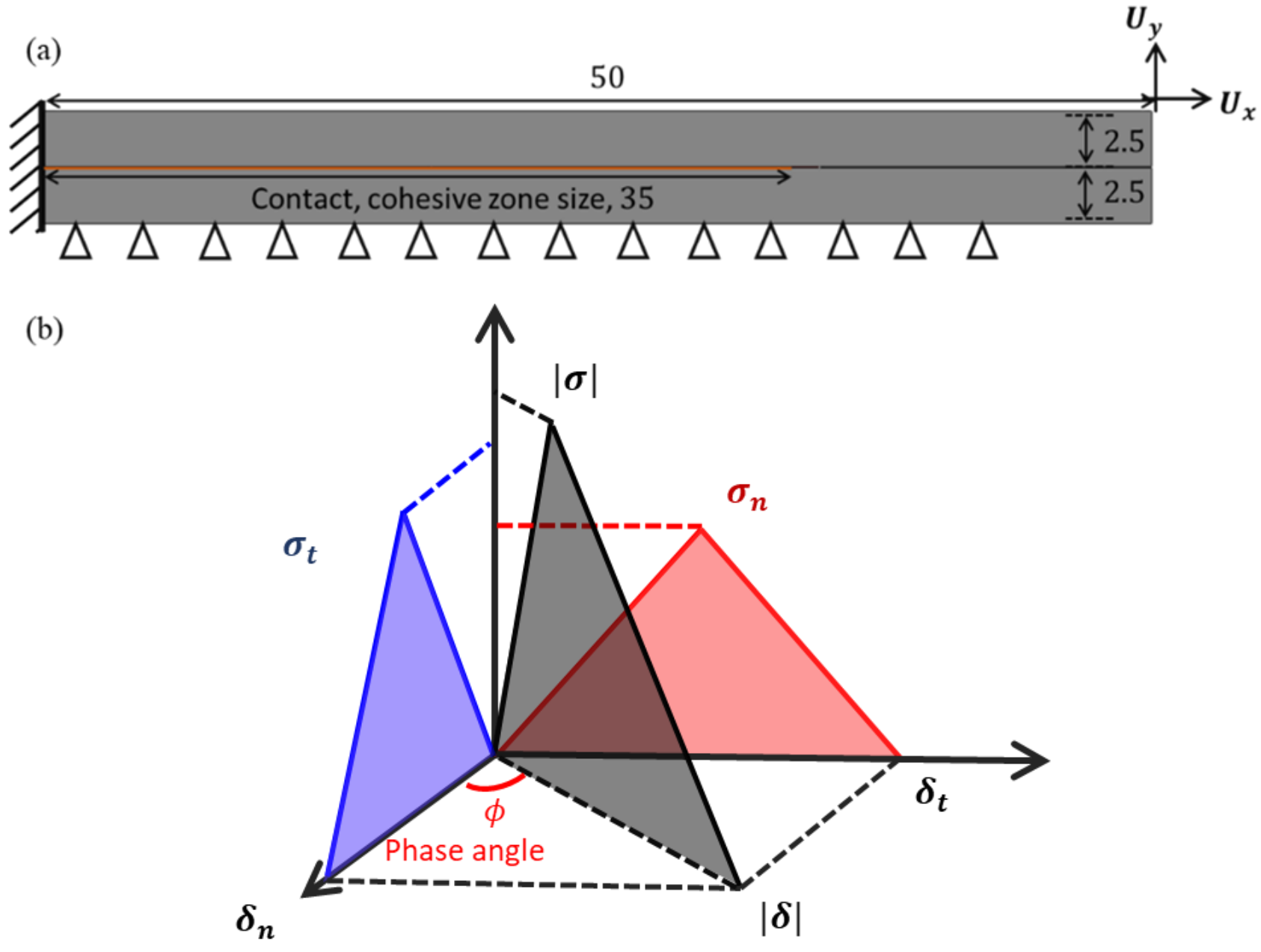

Figure 16. (a) End loaded split numerical model setup. (b) Traction-separation relations used for the interlayer.

For validating the TCNN model, each of the eleven paths (excluding the zero traction paths) is used in turn as the validation data set while the remaining ten paths are used for training. The TCNN prediction results for the loading path ($\phi = 14.30$) are shown in Figure 17. Note that, although proportional loading paths were selected based on the applied displacement, the paths that were actually followed by the cohesive element closest to the crack tip were not proportional [37]. The modeling results for the ten other cases are shown in Figures S23-32. The neural network experienced the greatest difficulty around the peak tractions when validation paths whose phase angles were less than 35° were used (Fig. S29-32). Alternatively, this essentially corresponds to training paths that were mode II dominant.

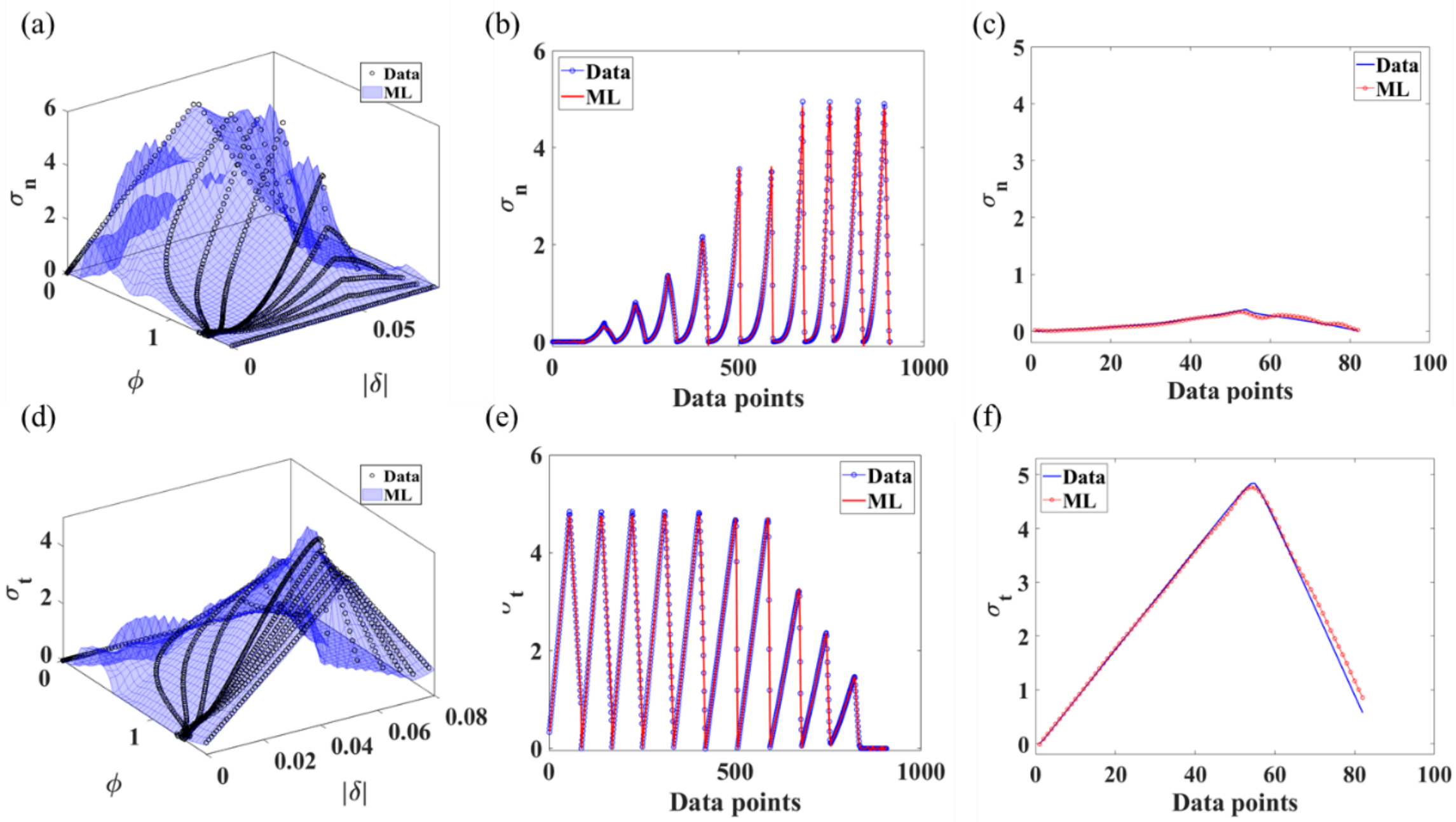

Figure 17. Sample modeling results for normal and tangential stresses obtained from the numerical model with 11 paths, where the loading path with $\phi = 85.22$ is used as validation data. (a) Predicted normal stress surface. Comparison between data and ML predicted results for normal stresses of (b) all data sets and (c) validation data sets. (d) Predicted tangential stress surface. Comparison between data and ML predicted results for tangential stresses of (e) all data sets and (f) validation data sets.

The prediction errors for each of the validation loading paths are summarized in Figure 18. The highest error for normal and tangential tractions were 9.49 % and 9.67 %, respectively. Nonetheless, the error in normal tractions was generally higher than that associated with shear tractions. The error in shear tractions was notably lower for validation paths that were mode II dominant. This may be due to the fact that the intervals between the training paths in these cases were smaller, thereby leading to training data that was closer to the validation data.

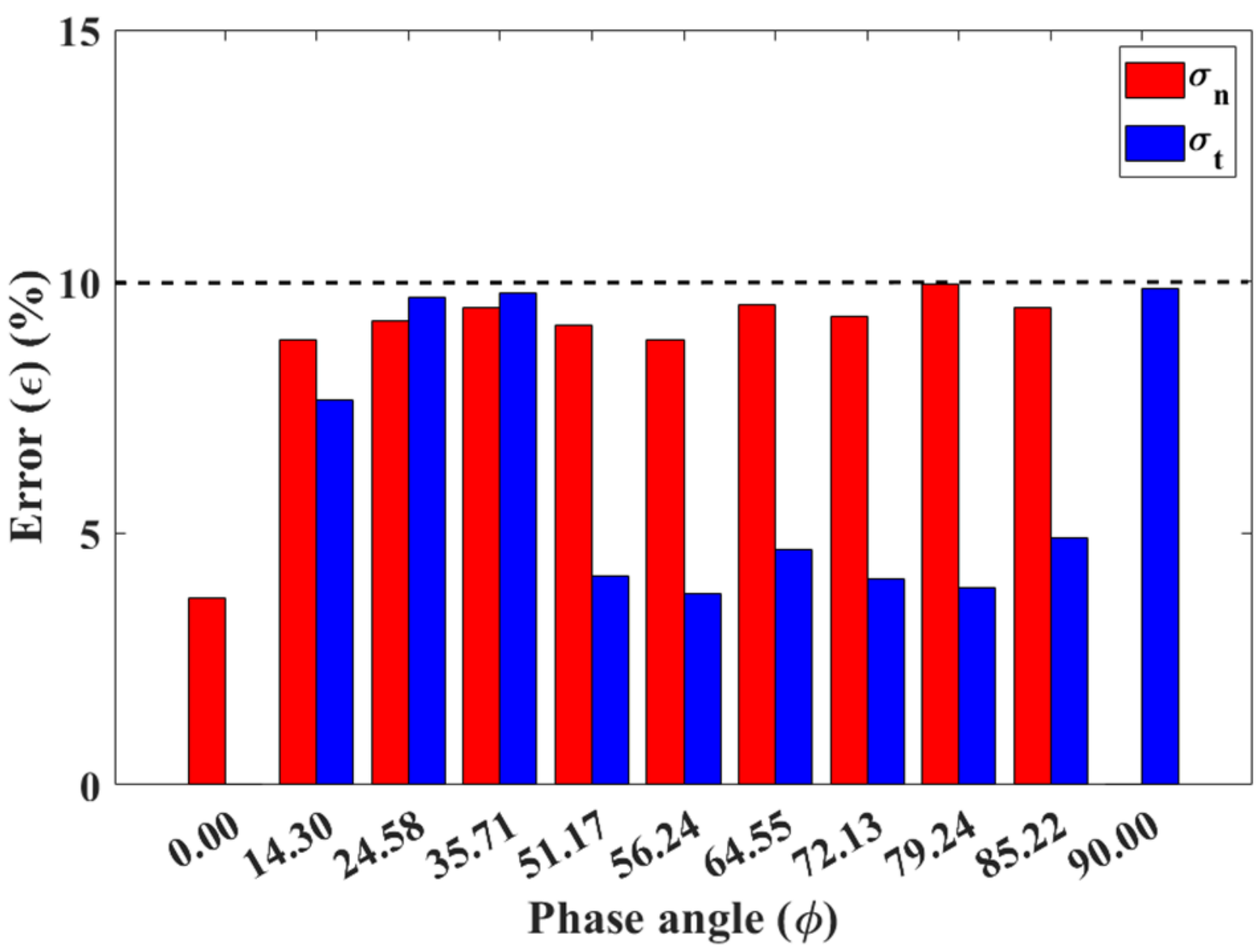

Figure 18. Prediction error for all loading paths.

### 4.3 Validation with experimentally generated data

Ten loading paths are included in the experimental data base that was considered here. Except for the $\phi = 87.5$ case, where only the tangential traction-separation relations are recorded, all other cases were mixed-mode ones. The number of data points for each loading path is noted in Table 2. The entire data base has a total of 1233 data points, including 593 and 640 data points for the normal and tangential components of the traction-separation relations, respectively. Considering the unique mode-mix dependent J-integral for the known Si-epoxy interface [37], the additional constrains have been implemented

Table. 2 Number of data points for each loading path

| **Phase angle (°)** | **27.0** | **-39.8** | **4.8** | **-53.1** | **35.3** | **61.1** | **87.5** | **18.1** | **-27.8** | **74.8** |
|---|---|---|---|---|---|---|---|---|---|---|
| **Number of data points** | **64** | **54** | **64** | **74** | **72** | **74** | **47** | **63** | **70** | **58** |

The data base was separated into training and validation data sets using two strategies: (1) Using each of the ten loading paths in turn as the validation data set while the remainder of the loading paths in the data base were used as the training data. This provides a broader perspective on the idea of using entire loading paths for data input and certainly removes some degree of bias by considering all the possibilities offered by the data base. (2) 10% of the experimental data was randomly selected across all loading paths as the validation data set while the remaining 90% was used for training. The ratio was preserved in considering five different cases of this scenario.

***Scheme 1: Loading paths***

The results obtained from the TCNN model (TCNN4) are compared (Figure 19) with a validation data set that was generated in the first approach for a loading path that followed a phase angle of 27.0°. In Figures 19a, d, respectively, the normal and tangential traction surfaces constructed using the TCNN model are presented alongside the input data. Greater scrutiny is provided (Figures 19 b-c, e-f) by comparing the validation data set with the tractions provided by the TCNN model along the 27.0° phase angle loading path. Using data along particular loading paths for training did not produce a very favorable result. In the nine other cases that are presented as Supplementary Information in Figures S33-41, some of the input data sets produced better agreement between the model results and the validation data, but some were worse.

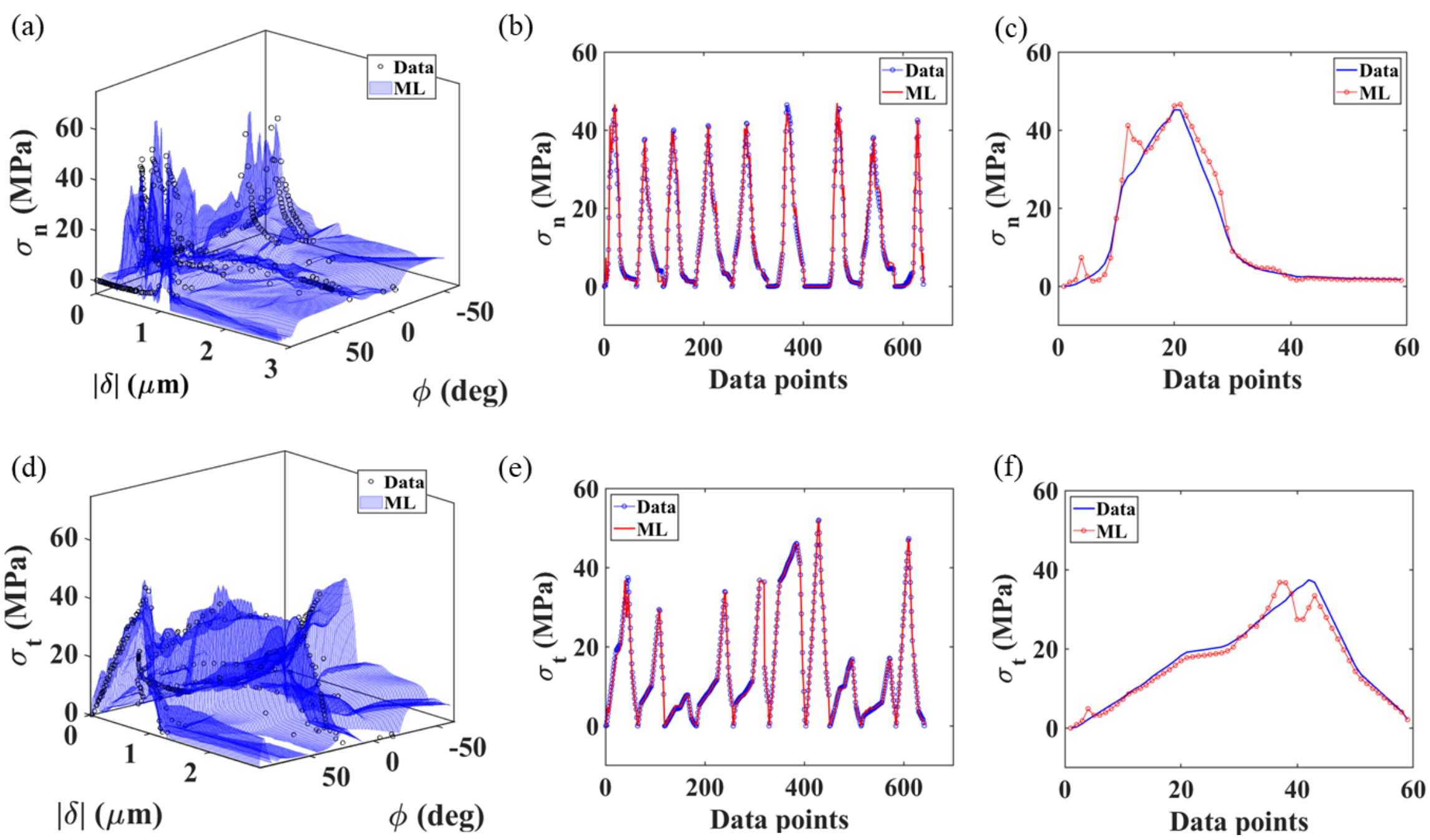

Figure 19. Comparing the normal and shear tractions produced by the TCNN where a loading path with a phase angle of 27.0° is used as the validation data set. Figures (a) and (d) present the input data and the normal and shear traction surfaces generated by the TCNN, while (b) and (e) compare results to all data sets and the (c) and (e) results to the validation data along the loading path.

The error between each of the validation data sets and the results produced by the TCNN model is expressed as

$$\epsilon = \frac{\sum_{i=1}^{n}(Y_i^l - Y_i)}{\sum_{i=1}^{n}(Y_i)}, \qquad (21)$$

where $Y_i$ is the $i^{\text{th}}$ traction value of the validation data sets and the $Y_i^l$ is the value obtained from the TCNN model. The quantity $n$ is the total number of data points in a particular validation data set. Across all the input data sets, the error ranged from 2.86 to 9.87 % (Figure 20). The input data set which used the loading path with a phase angle of -39.8 as the validation data set produced the lowest error in the normal tractions while the 27.0 loading path had the lowest error in the tangential tractions.

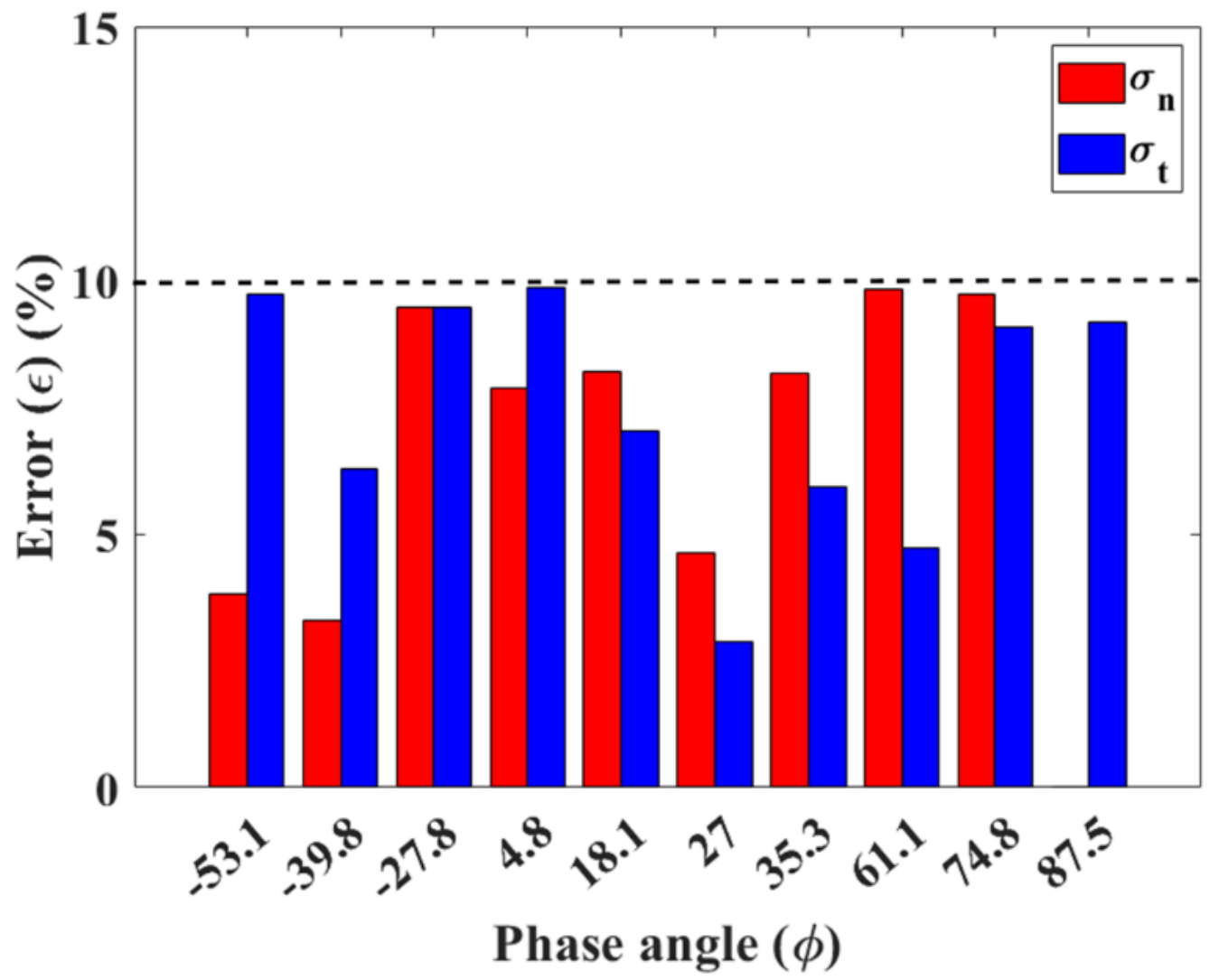

Figure 20. Difference between the TCNN results and the validation data for each of the loading paths that were considered.

***Scheme 2: Randomly selected points***

The modeling results for validation data that was randomly selected from 10% of the entire data set are shown in Figure 21. It can be seen (Figure 21a, d) that the TCNN was able to generate the normal and tangential traction surfaces. The data points that were used to generate the surfaces also appear there. A comparison of the modeling results with the validation data set is provided in Figure 19 b-c, e-f. While there are some random outliers, the TCNN results are generally in quite reasonable agreement with the validation data. The same comparisons are made for the other four input and validation data sets (Figures. S42-S45).

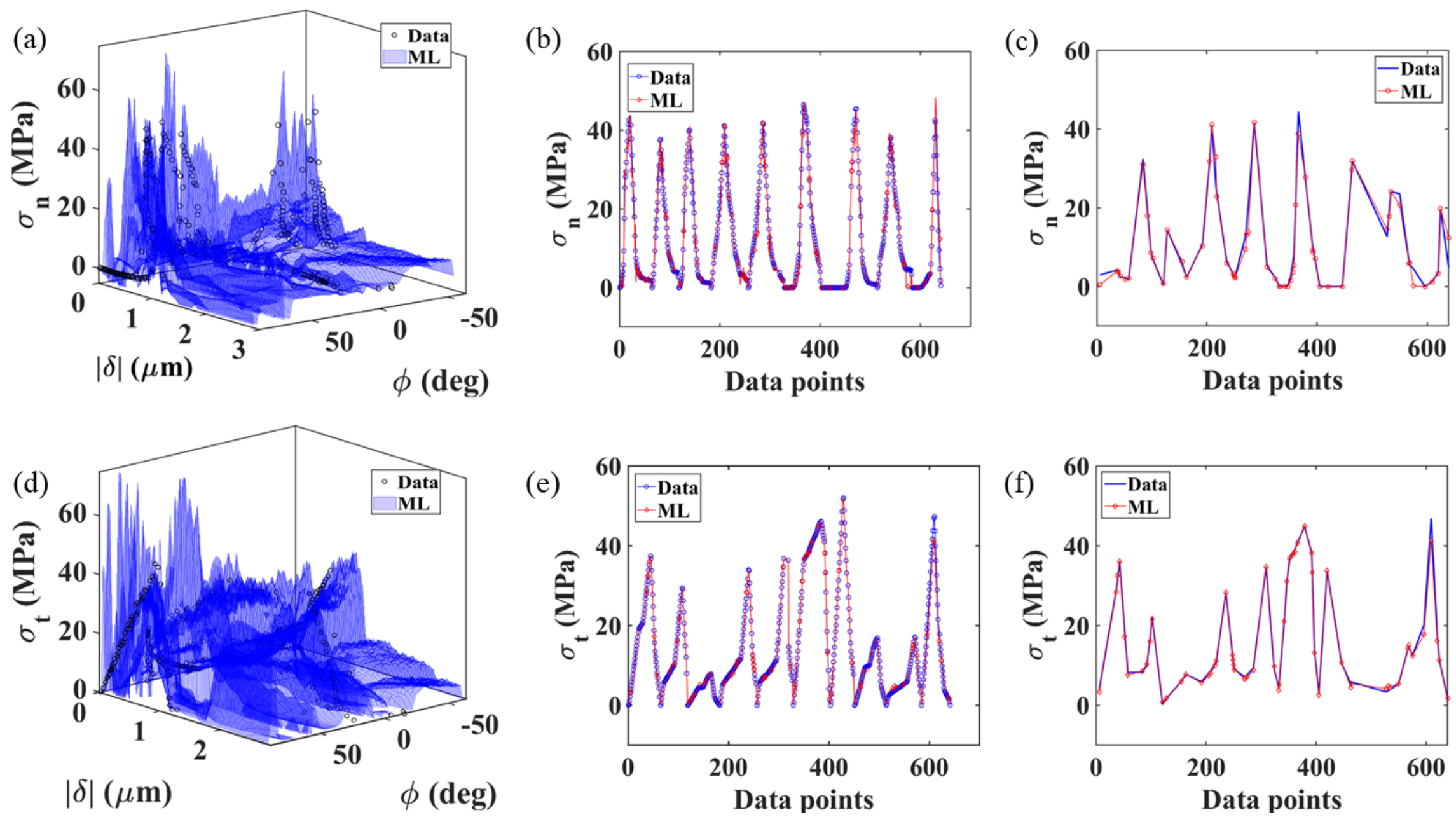

Figure 21. Comparing the normal and shear stresses produced by the TCNN where a loading path 10% of the entire data set is used as the validation data set (case 1). Figures (a) and (d) present the input data and the normal and shear traction surfaces generated by the TCNN, while (b) and (e) compare the TCNN results to all data sets and (c), (f) to the validation data.

The results just presented made use of 90% of the entire data set for training the TCNN. The next step that was considered was to reduce the amount of data that was used for training purposes. Training data sets that consisted of 80, 70, 60 and 50% of the entire data set were evaluated. For each fraction of input data, five trials that made use of randomly different data points were contemplated. The results of this exercise are summarized in the error plot that appears in Figure 22. Within each fraction of data points in the validation set (complement of the training data set), the data points joined by the connecting lines reflect the average error for the selected fraction of the entire data that was used. The error bars represent the maximum and minimum errors for each ratio. The results show that the error between the results produced by the TCNN, and the validation data increases monotonically as the fraction of data points that are used in training decreases, which is not unexpected. Once again, the error in the normal tractions was slightly lower than the shear tractions.

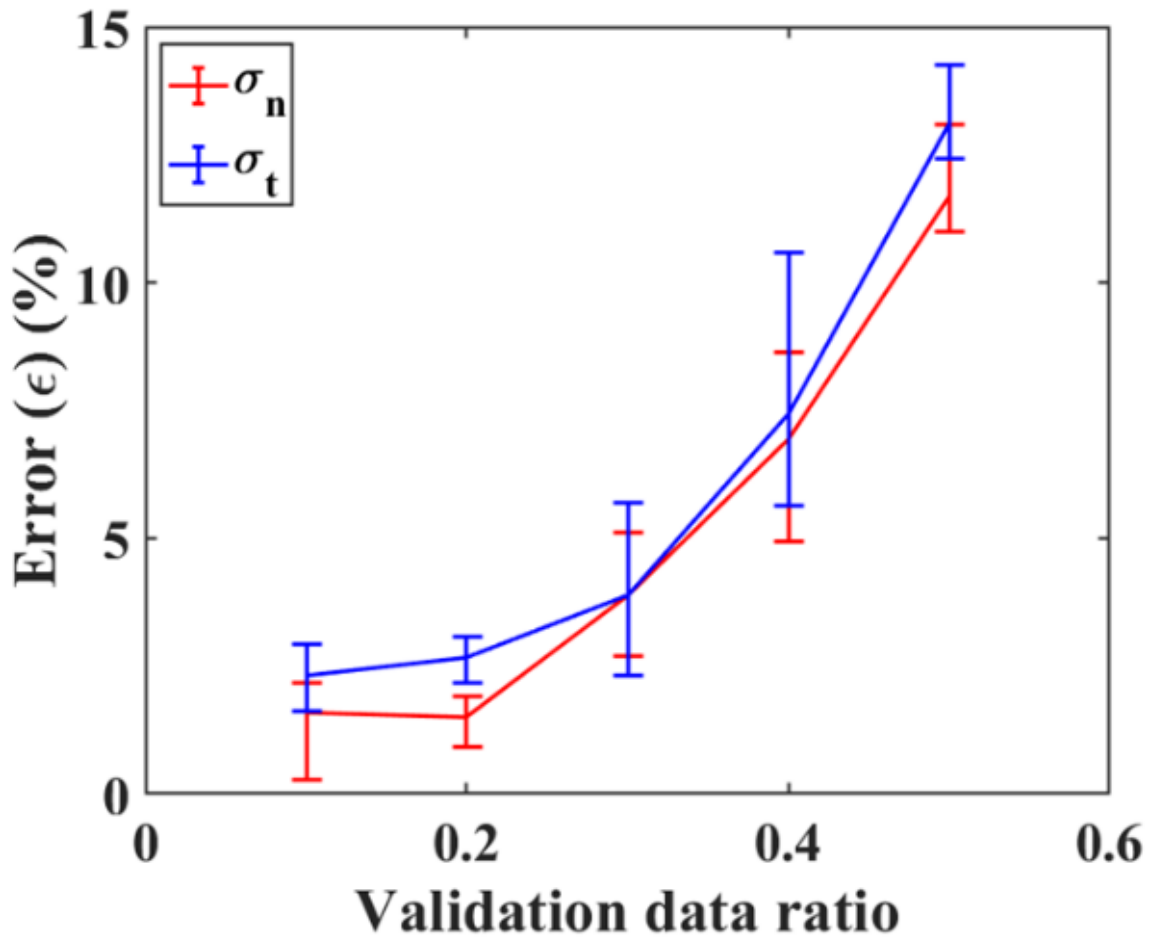

Figure 22. Average difference between the TCNN results and the validation data for smaller input data sets. Five data sets were considered for each ratio of the number of data points used for validation or training and the error bars reflect the maximum and minimum errors.

**4.3 Hyperparameter study**

The results that have been presented to date were developed with a neural network that consisted of two layers and 60 neurons per layer. We now consider the effect of the structure of the TCNN on the development of traction surfaces by allowing for one, two or three layers with 20 to 100 neurons per layer in increments of 20. Models with and without all three thermodynamic constraints are studied along for a total of 30 cases. For all cases, the learning rate is fixed at $3 \times 10^{-3}$ and the number epochs for training is fixed at 100,000. This ensures the convergence of all cases. An example of the convergence based on the total loss is illustrated (Figure S46)

The final loss and the difference between the model results and validation data for all the cases that were run without thermodynamically consistent constraints appear in Figure 23 using the dataset used for Case 1. Its counterpart with the first three constraints activated follows in Figure 24. In both situations, the final loss decreases rapidly between one and two layers. However, the number of neurons has a limited influence on the convergence when thermodynamic constraints are not applied. With and without constraints, the difference between the model results and the validation data is relatively high for over-simplified structures (e.g., one layer with 20 neurons). It decreases as the number of layers is increased, but overfitting is observed, as indicated by the slight increase in the error associated with the modeling, when there are more than 60 neurons per layer. This is particularly true when no thermodynamic constraints are applied. For all cases with or without thermodynamically consistent constraints, the structure with two layers, 60 neurons each layer could produce prediction errors that are less than 12% and outruns other structural setups, which either have problems of bad convergence or overfitting problems.

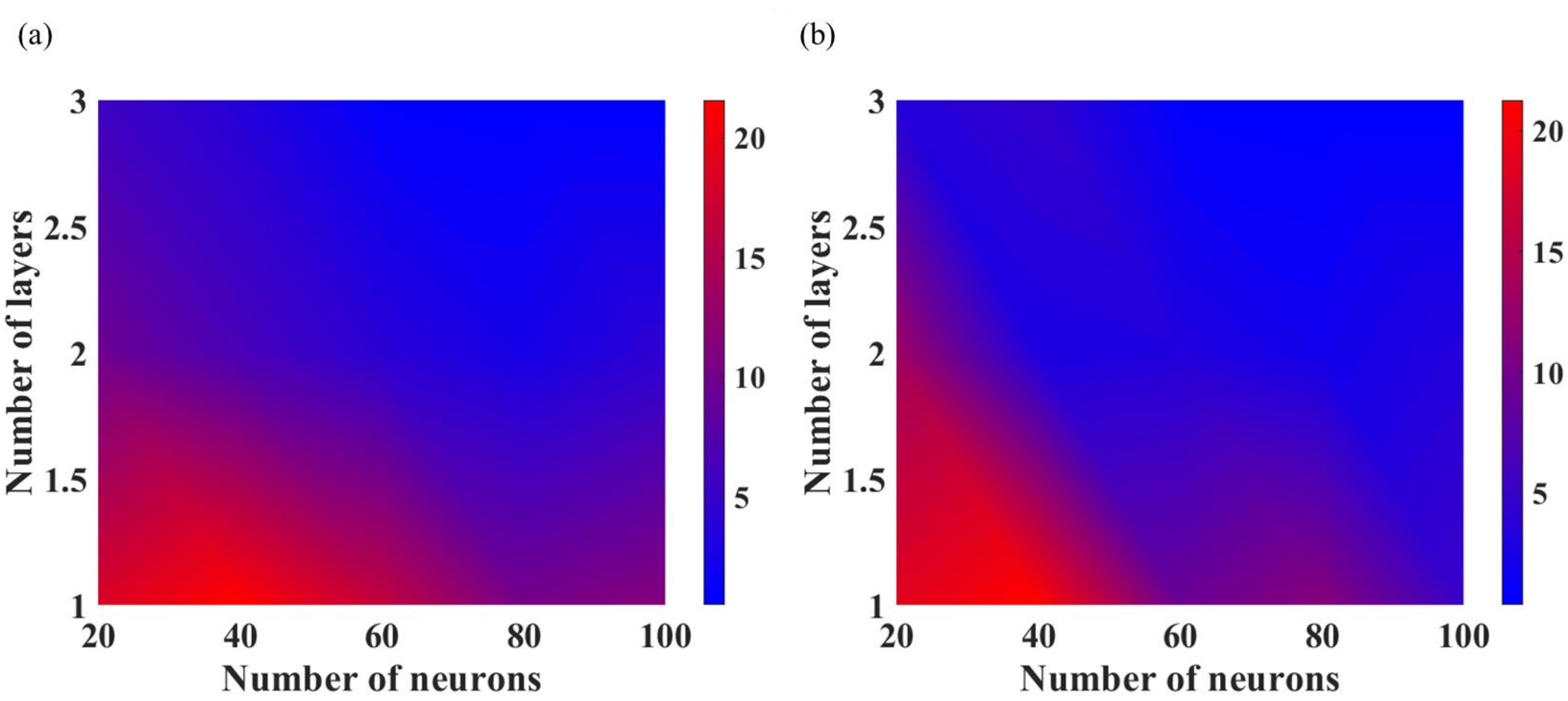

Figure 23. TCNN without thermodynamic constraints (Case 3). (a) Total loss. (b) Prediction error.

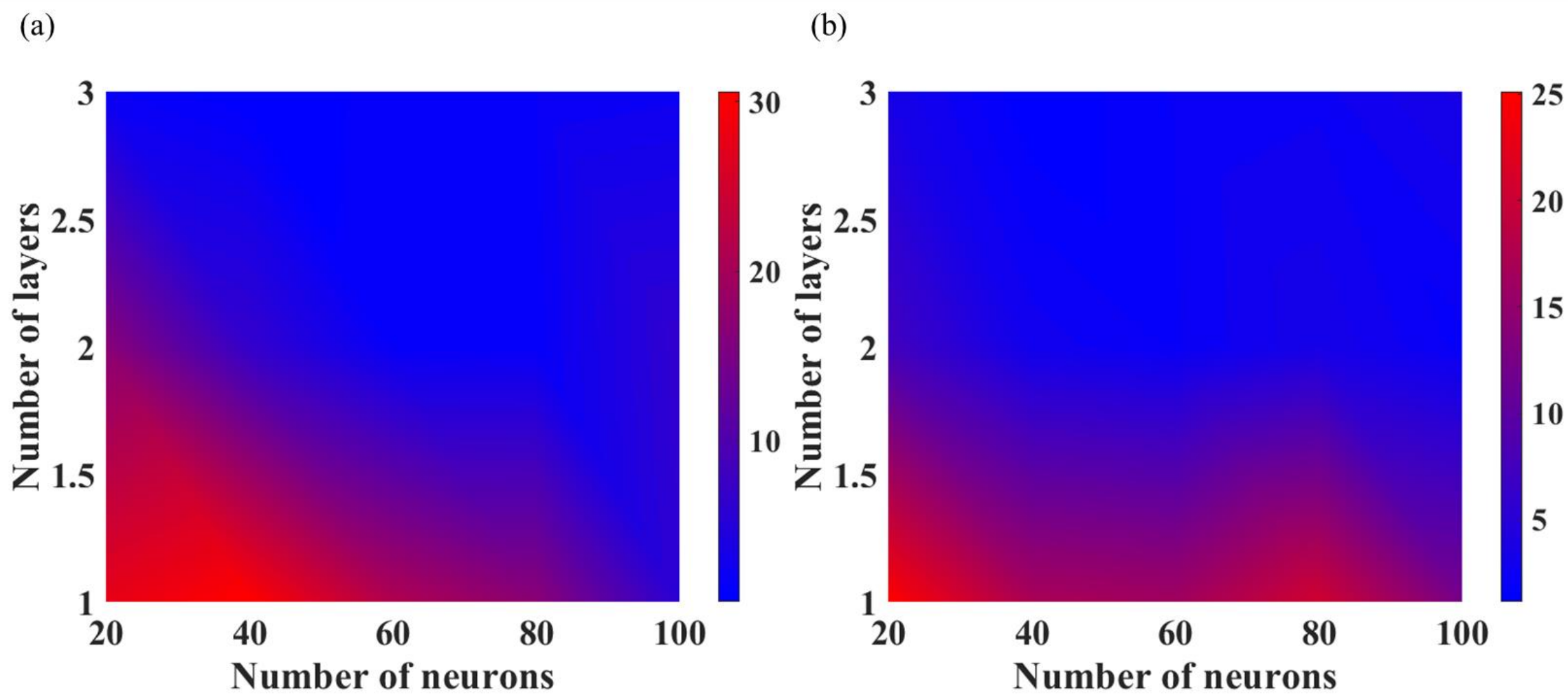

Figure 24. TCNN with all four thermodynamic constraints (Case 3). (a) Total loss. (b) Prediction error.

## 5. Results and discussion

As a basis for discussion of the results obtained from the ideas presented above, we first obtained the surfaces for mixed-mode traction-separation relations using a deep neural network without any constraints. Subsequently, we discuss the improvements that can be obtained by adding thermodynamic constraints with three sets of selected weight factors and then demonstrate the effectiveness of the weight factors selected by the Bayesian optimization algorithm. The results shown in this section are trained based on whole experimental data sets sparsely distributed in traction-separation space.

### 5.1 Deep neural network

The normal and tangential traction-separation relation surfaces obtained from only using the basic deep neural network are illustrated (Figure 25a-b) with respect to $|\delta|$ and $\phi$. The traction-separation relations that are embodied in the result were extracted and are compared (Figure 25c-d) with the input data sets along the various loading paths that were considered in the experiments. The agreement was quite reasonable.

We then evaluated how thermodynamically consistent the result was. Based on the surfaces for traction-separation relations that were obtained with thermodynamic constraints incorporated in the deep neural network, violations to each of the three thermodynamically consistent conditions are calculated separately as follows.

For the TC1 condition, we introduce the violation term, ($\mathrm{Vio}_1$), which represents negative energy dissipation and was calculated based on gradient of the J-integrals calculated from the predicted values of the tractions from

$$\mathrm{Vio}_1 = \left|\min\left(\frac{\partial J_n}{\partial|\delta|}, 0\right)\right| + \left|\min\left(\frac{\partial J_t}{\partial|\delta|}, 0\right)\right|. \tag{23}$$

This equation is used to identify data points that have a negative gradient of the J-integral components. Their normalized values ($\overline{\mathrm{Vio}}_1$), with respect to the maximum value

$$\overline{\mathrm{Vio}}_1 = \frac{\mathrm{Vio}_1}{\max\left(\mathrm{Vio}_1\right)} \tag{24}$$

are illustrated as the red contour in Figure 25e.

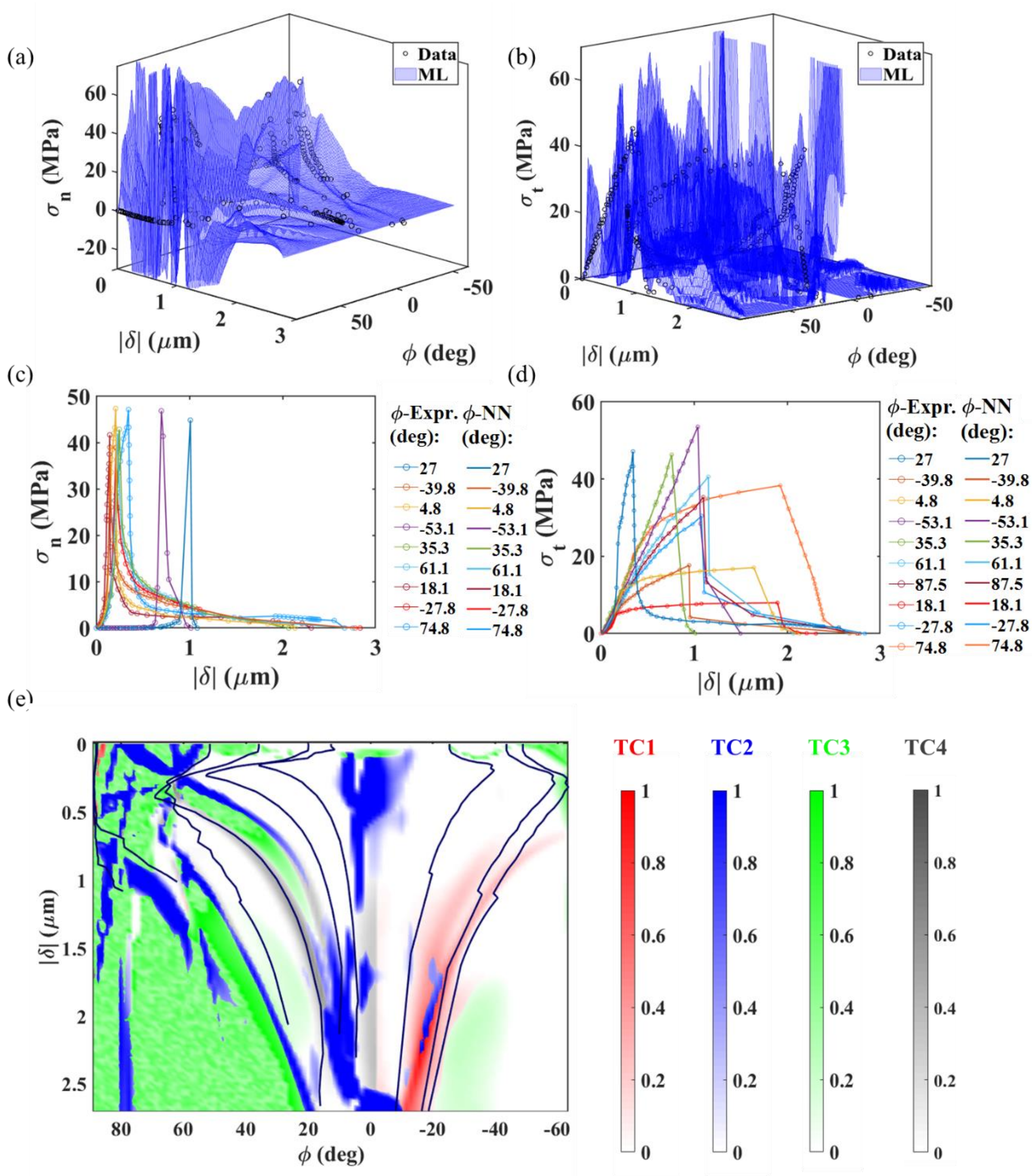

Figure 25. (a) Normal and (b) tangential traction surfaces obtained using the deep neural network by itself without any constraints (TCNN0). Comparison with (c) normal and (d) tangential tractions obtained experimentally at specific mode-mixes. (e) Violations associated with each of the four thermodynamically consistent conditions: The red, blue, green and grey contours represent the degrees of violation of conditions TC1, TC2, TC3 and TC4 respectively, while the black lines are the loading paths that were taken in the experiments.

For the TC2 condition, the damage surfaces are first calculated automatically based on predicted traction-separation relations for each iteration (defined in Equation 2). The gradient change is then calculated with respect to loading angle, $\theta$, for each point of the damage surface as shown in Figures 5b-c. Data points with maximum gradient decreases that are not along the $\theta = 0°$ direction are located and the violation ($\text{Vio}_2$) is calculated as the difference between the fastest descending direction and 0°, as shown in following equation.

$$\text{Vio}_2 = |\theta_{ddn}| + |\theta_{ddt}|, \tag{25}$$

where $\theta_{ddn}$ and $\theta_{ddt}$ are the angle of fastest descending direction in the normal and tangential traction-separation relation surfaces. The normalized value ($\overline{\text{Vio}_2}$) is taken to be

$$\overline{\text{Vio}_2} = \frac{\text{Vio}_2}{\max(\text{Vio}_2)} \tag{26}$$

with respect to maximum value and is shown as the blue contour in Figure 26e.

For the third condition (TC3), the ratio between the normal and tangential tractions ($\sigma_n, \sigma_t$) is calculated and compared with the tangent of the phase angle, $\phi$. Noting that exact equivalence is usually impossible to achieve, we use a threshold angle ($\epsilon_{phi}$) of 5° and violation ($\text{Vio}_3$) is reached when the difference between the ratio and the phase angle tangent is larger than $\epsilon_{phi}$: Thus

$$\text{Vio}_3 = \max\left(\left|\frac{\sigma_t}{\sigma_n} - \tan\phi\right|, \tan(\epsilon_{phi})\right) - \tan(\epsilon_{phi}) \tag{27}$$

Again, a normalized value ($\overline{\text{Vio}_3}$)

$$\overline{\text{Vio}_3} = \frac{\text{Vio}_3}{\max(\text{Vio}_3)}, \tag{28}$$

is introduced and then depicted as the green contour shown in Figure 25e.

For the final condition (TC4), the violation is defined as the toughness change contradicting the monotonicity along phase angle direction. Thus,

$$\mathrm{Vio}_4 = \max\left(\frac{\partial J_n}{\partial|\phi|}, 0\right) + \max\left(-\frac{\partial J_t}{\partial|\phi|}, 0\right), \text{ for } |\delta| = \left|\delta_f\right| \quad (29)$$

Like before, a normalized value ($\overline{\mathrm{Vio}_4}$)

$$\overline{\mathrm{Vio}_4} = \frac{\mathrm{Vio}_4}{\max\left(\mathrm{Vio}_4\right)}, \quad (30)$$

is calculated and shown as the grey contour in Figure 25e.

It can be seen that a significant portion of the violation map (Figure 25e) is colored, indicating that there is a substantial degree of violation in the traction-separation relation surfaces that are produced by the deep neural network operating in the absence of any constraints. The strongest thermodynamic consistency condition (TC1, red) was violated over relatively small portions of the surfaces. There are a lot of blue and green regions in the map corresponding to areas on the surfaces where the second and third consistency conditions are violated. In addition, a number of the loading paths taken in the experiments pass through these regions, suggesting that enforcing this condition in the learning process will be helpful. Finally, the region where the fourth condition is violated is relatively small, making this a candidate for a relatively low weight factor.

## 5.2 Thermodynamically consistent deep neural network (TCNN)

The effect of the different weight factors (Table 1) that were applied to each of the thermodynamically consistency conditions is illustrated in Figures 26-29, corresponding to TCNN1 to TCNN4, respectively.

What can be learned by comparing the results from Figures 26-29? First, it can be seen (sub-Figures c-d) that the surfaces produced by all four cases that were considered here compared well with the input data from the experiments. Thus, the main features of the input data are captured in the training process. Shifting our focus to the degree of violation (sub-Figure e) of each of the conditions, adding the constraints reduced the amount of violation compared to the simple neural network (Figure 25e). Furthermore, the amount of overlap between violations decreased as more constraints were added. Nonetheless, even for the best combination of weight factors considered here (TCNN4), portions of the loading paths were still in violation. This suggests that systematic errors and noise in the experimental data lead to violations and provides an opportunity to improve certain aspects of the data, particularly if the main contributors can be identified. Another thing to consider is that the examples provided in this section are only four of a potentially infinite number of combinations of weight factors. The next section presents the results of the Bayesian approach (Section 3.3) to optimize the selection of weight factors.

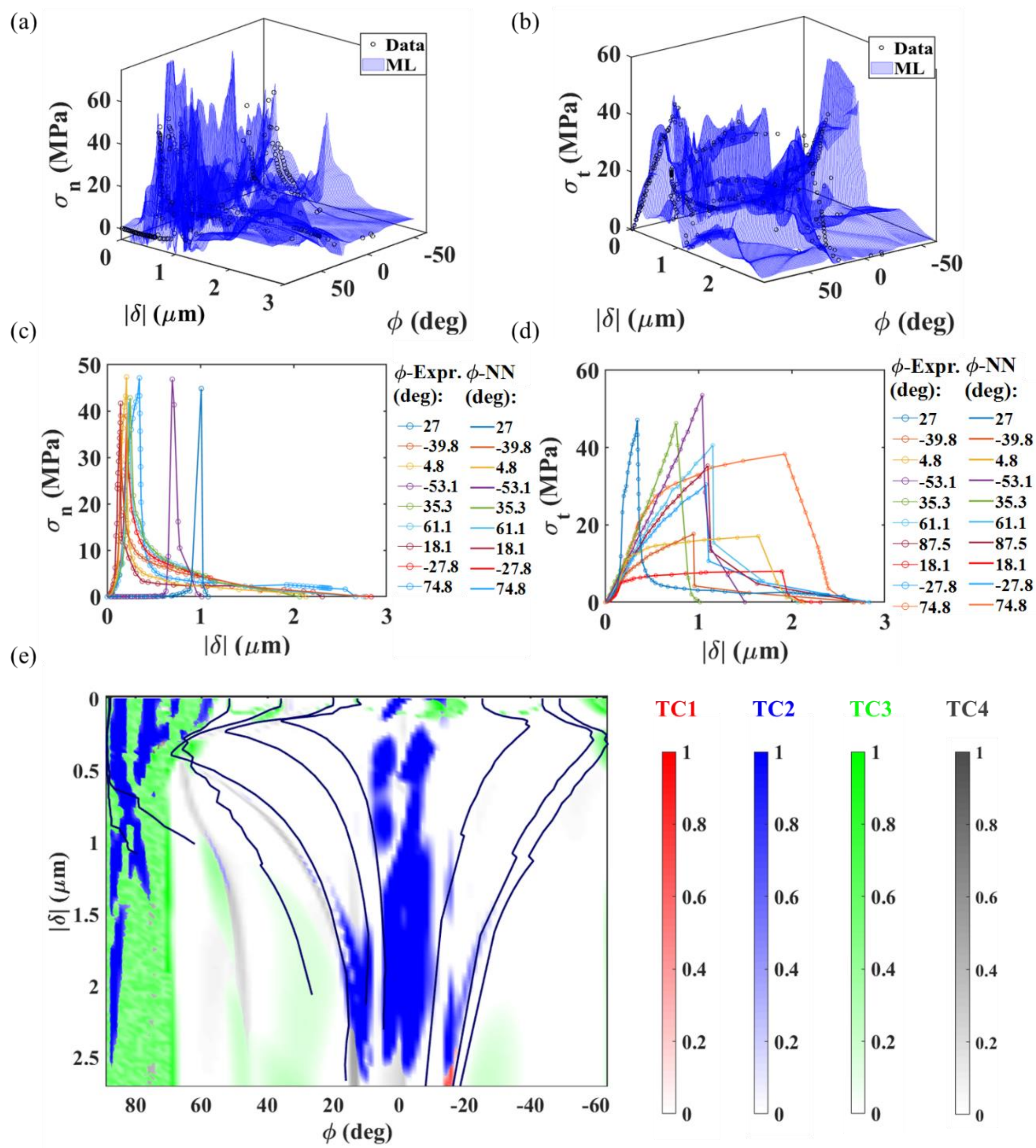

Figure 26. Results for the TCNN1 operating with MSE ($\lambda_0 = 0.9$) and only the TC1 constraint ($\lambda_1 = 0.1$): Surfaces for (a) normal and (b) tangential tractions; comparing results for (c) normal and (d) tangential tractions with those from experiments. (e) Violations associated with each of the four thermodynamically consistent conditions: The red, blue, green and grey contours represent the degrees of violation of conditions TC1, TC2, TC3 and TC4 respectively, while the black lines are the loading paths that were taken in the experiments.

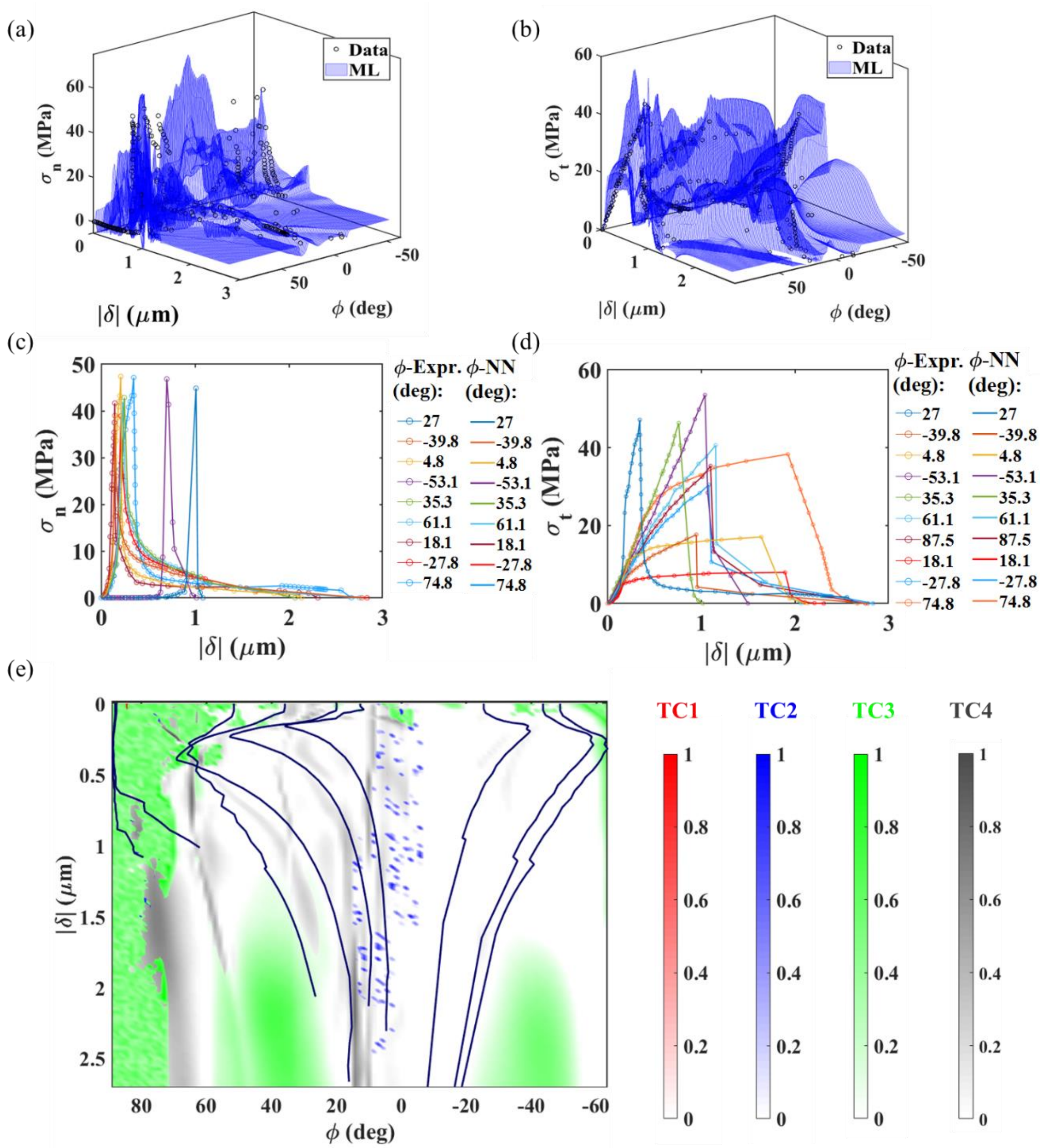

Figure 27. Results for the TCNN2 operating with MSE ($\lambda_0 = 0.8$) and the conditions TC1 and TC2 ($\lambda_1 = 0.1, \lambda_2 = 0.1$): Surfaces for (a) normal and (b) tangential tractions; comparing results for (c) normal and (d) tangential tractions with those from experiments. (e) Violations associated with each of the four thermodynamically consistent conditions: The red, blue, green and grey contours represent the degrees of violation of conditions TC1, TC2, TC3 and TC4 respectively, while the black lines are the loading paths that were taken in the experiments.

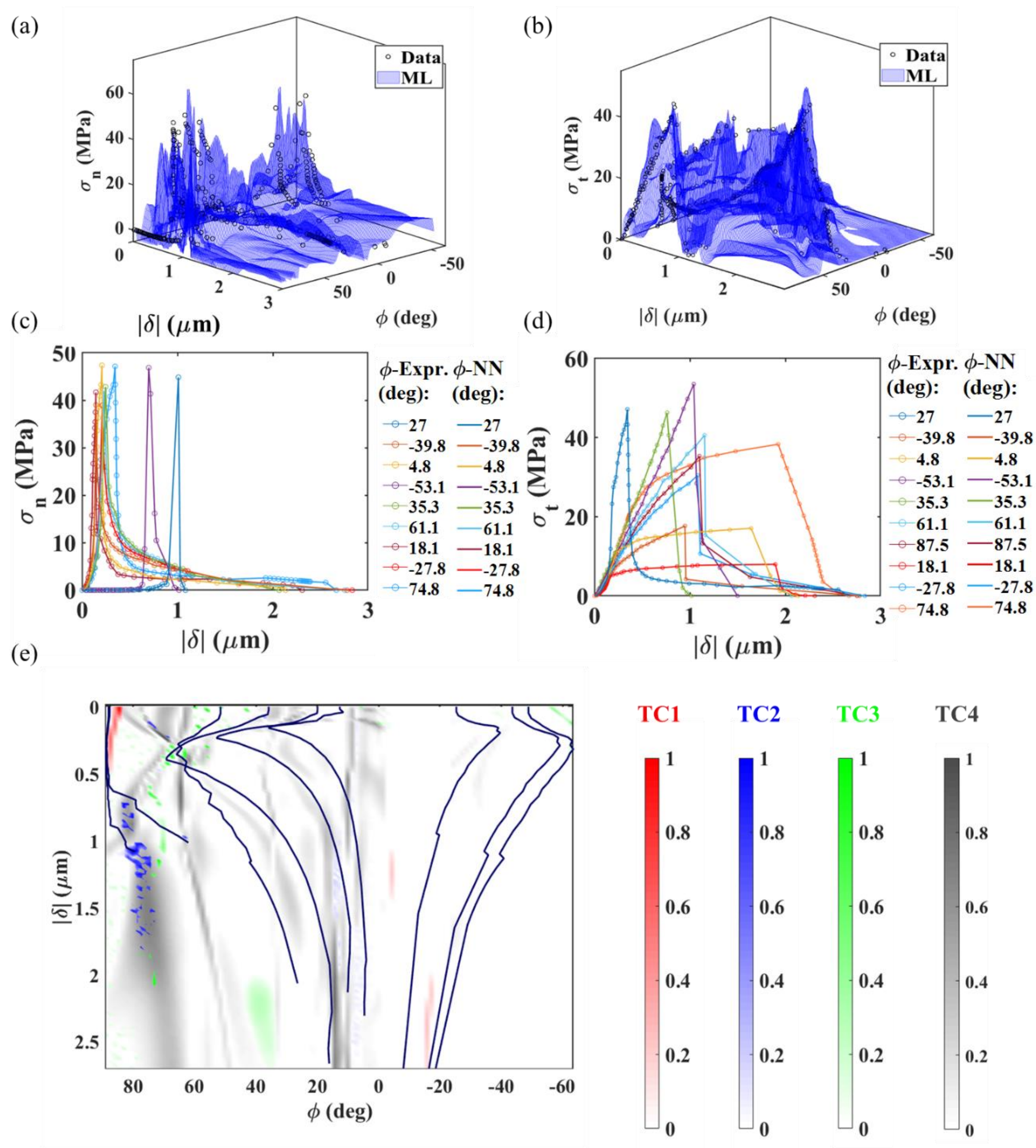

Figure 28. Results for the TCNN3 operating with MSE ($\lambda_0 = 0.78$) and the conditions TC1, TC2 and TC3 ($\lambda_1 = 0.1, \lambda_2 = 0.1, \lambda_3 = 0.02$): Surfaces for (a) normal and (b) tangential tractions; comparing results for (c) normal and (d) tangential tractions with those from experiments. (e) Violations associated with each of the four thermodynamically consistent conditions: The red, blue, green and grey contours represent the degrees of violation of conditions TC1, TC2, TC3 and TC4 respectively, while the black lines are the loading paths that were taken in the experiments.

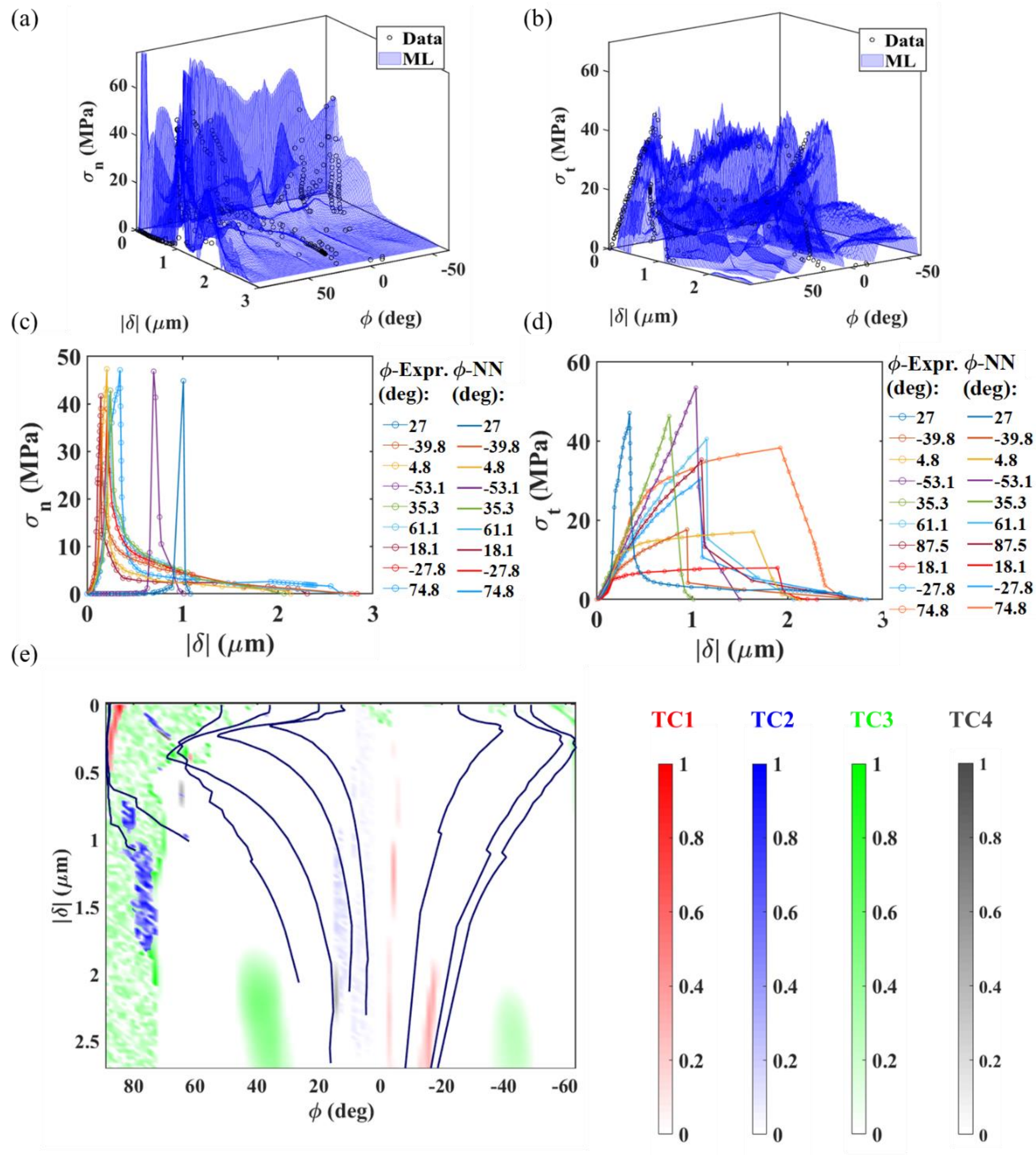

Figure 29. Results for the TCNN4 operating with MSE ($\lambda_0 = 0.68$) and the conditions TC1, TC2, TC3 and TC4 ($\lambda_1 = 0.1, \lambda_2 = 0.1, \lambda_3 = 0.02, \lambda_4 = 0.1$) : Surfaces for (a) normal and (b) tangential tractions; comparing results for (c) normal and (d) tangential tractions with those from experiments. (e) Violations associated with each of the four thermodynamically consistent conditions: The red, blue, green and grey contours represent the degrees of violation of conditions TC1, TC2, TC3 and TC4 respectively, while the black lines are the loading paths that were taken in the experiments.

## 5.3 Bayesian optimization of weight factors

Based on the Bayesian optimization algorithm proposed in Section 3.3, the weight factors for loss function terms corresponding to the MSE as well as the constraints (TC1-TC4) are optimized. The convergence of the process is shown in Figure 30 after 1000 iterations were conducted. The value of the loss function initially decreases continuously and reduces to 1.0 within 400 iterations. Ranges for each of the five weight factors ($\lambda_0 - \lambda_4$) and their optimized values are presented in Table 3. The largest importance is still awarded to the data from the experiments (MSE, $\lambda_0 = 0.501$). Second priority was given to the second law of thermodynamics (TC1, $\lambda_1 = 0.204$) and it was obeyed by all the input data sets. The weight factor associated with the second thermodynamic constraint was close ($\lambda_2 = 0.102$ ) to the value (0.1) that was used in the three examples described earlier. The optimized weight factor for the third condition (TC3, $\lambda_3 = 0.012$) is smaller than the setting that was used in TCNN3 settings, maintaining the position of this constraint as the lowest in importance. The fourth condition (TC4, $\lambda_3 = 0.181$) is optimized with a larger value than the original setting. This indicates the importance of including the variation of toughness with mode-mix as a constraint in the training process.

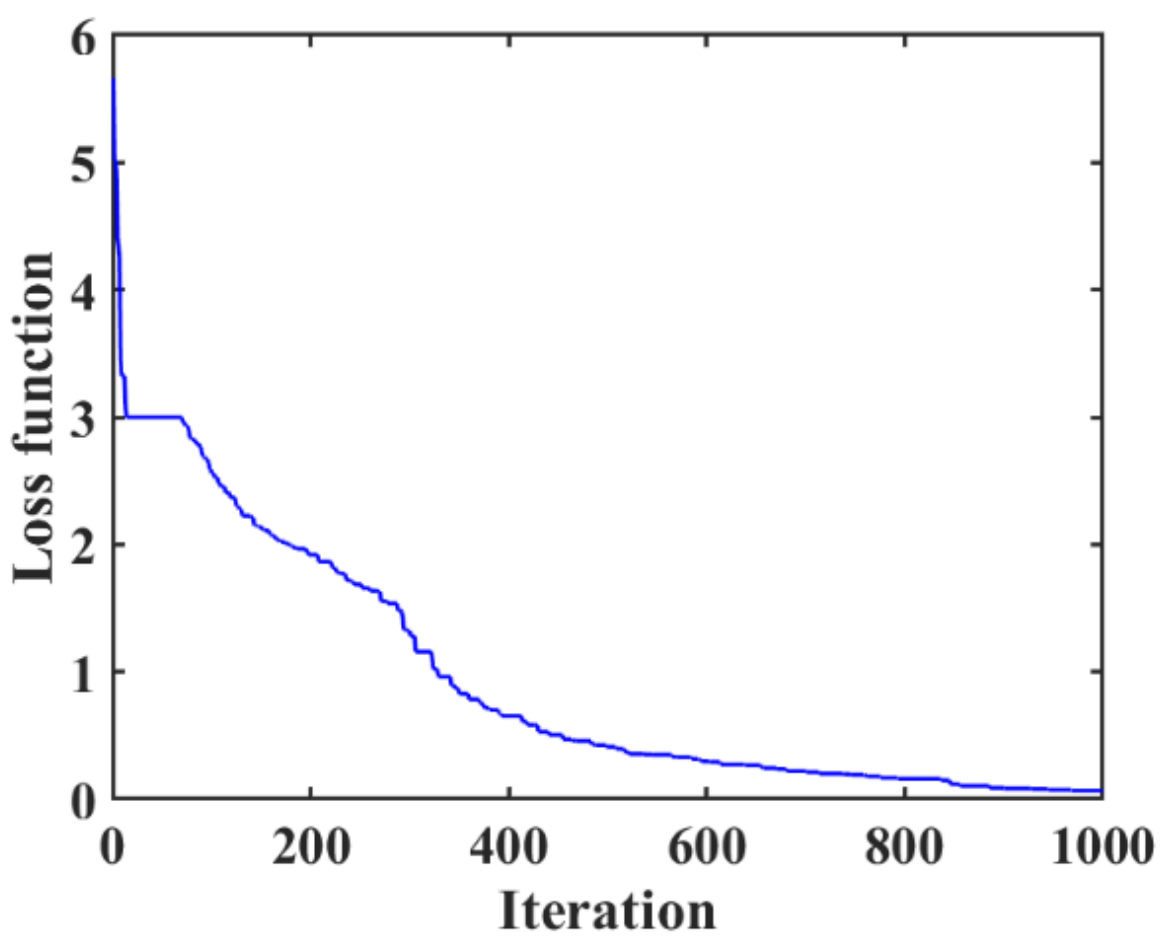

Figure 30. Convergence of the Bayesian optimization process for optimizing the values of the weight factors.

Table. 3 Optimized Weight Factors

| | Upper bound | Lower bound | Optimized |
|---|---|---|---|
| $\lambda_0$ (MSE) | 0.750 | 0.500 | 0.501 |
| $\lambda_1$ (TC1) | $(1-\lambda_0)\times 0.900$ | $(1-\lambda_0)\times 0.100$ | 0.204 |
| $\lambda_2$ (TC2) | $(1-\lambda_0-\lambda_1)\times 0.900$ | $(1-\lambda_0-\lambda_1)\times 0.100$ | 0.102 |
| $\lambda_3$ (TC3) | $(1-\lambda_0-\lambda_1-\lambda_2)\times 0.900$ | $(1-\lambda_0-\lambda_1-\lambda_2)\times 0.100$ | 0.012 |
| $\lambda_4$ (TC4) | NA | NA | 0.181 |

We used Figures 26-29 to illustrate the effect of various combinations of weight factors, so it is now instructive to return to that format in Figure 31 to observe the effect of the optimized selection of weight factors. The resultant traction surfaces are shown in Figure 31a-b. These are the surfaces that would be used in cohesive zone modelling of crack initiation and growth in structures such as microelectronics packages that contain silicon/epoxy (e.g., molding compound) interfaces. The veracity of these surfaces is first observed in Figure 31c-d, where the optimized neural network approach maintains good correspondence with the input experimental data even though the MSE weight factor was reduced to 0.501 from 0.680, the lowest value that was used in our initial trials.

Compared to the results shown in Figure 29e, Bayesian optimization, which also applied all four constraints, significantly reduced the degree of violation (Figure 31e). Furthermore, now only very small portions of the loading paths that were taken in the experiments are in violation of any of the four constraints. Thus, Bayesian optimization has effectively reduced the degree of conflict between the thermodynamic constraints and the input data. The trend of the experimental data is still fully captured while sufficiently honoring all four constraints. Bayesian optimization also confirms that the third constraint (TC3) is the least important one.

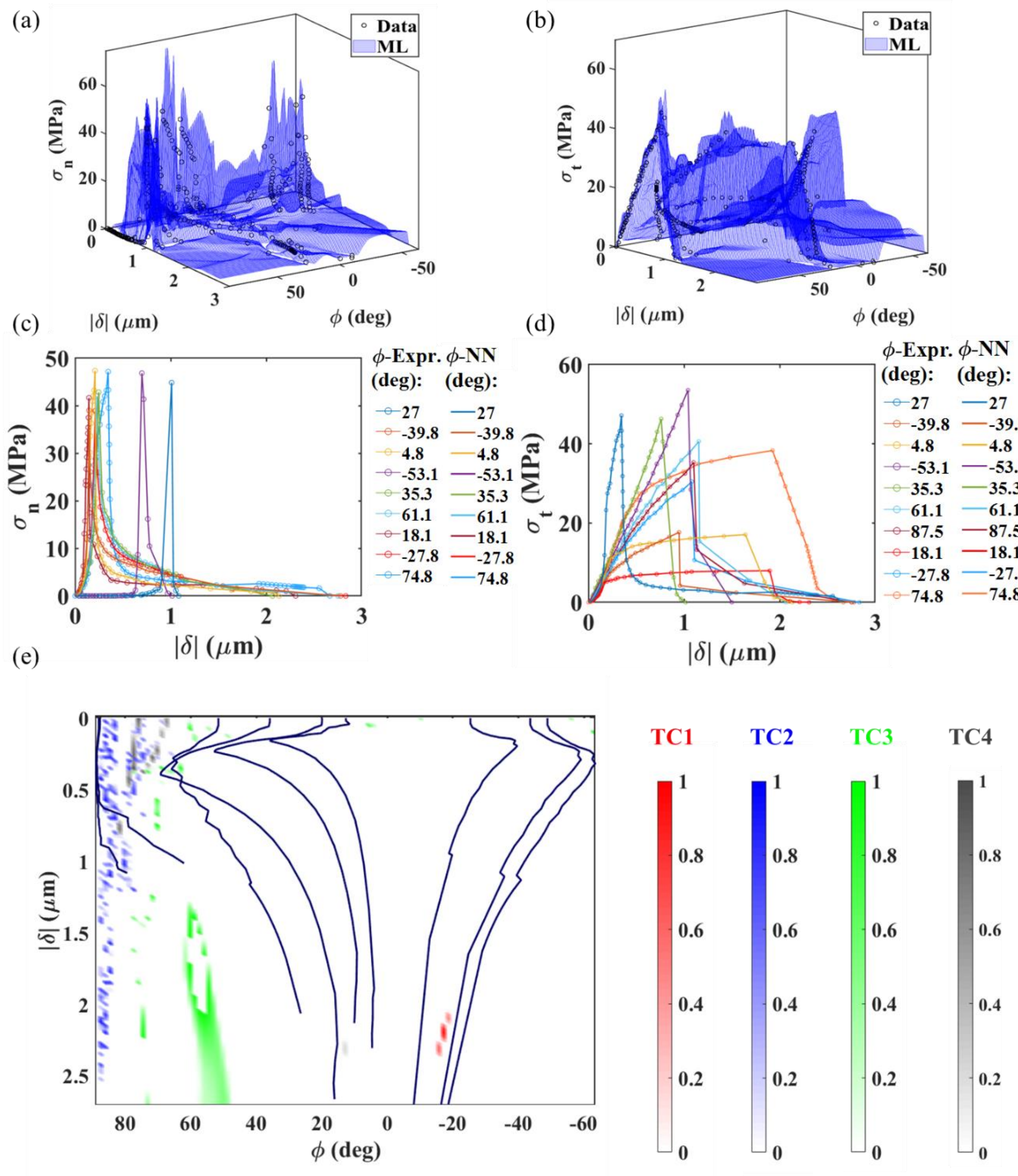

Figure 31. Results for the TCNN with Bayesian optimization. Surfaces for (a) normal and (b) tangential tractions; comparing results for (c) normal and (d) tangential tractions with those from experiments. (e) Violations associated with each of the four thermodynamically consistent

conditions: The red, blue, green and grey contours represent the degrees of violation of conditions TC1, TC2, TC3 and TC4 respectively, while the black lines are the loading paths that were taken in the experiments.

### 5.4 Prediction errors and constraint violations

The evaluation of the error and degree of violation of constraints that was presented in Figures 26-29c-e and 31c-e is made more quantitative by introducing the prediction error and violation rate. The prediction error is defined as:

$$\text{Prediction error} = \text{Es}\left(\left\|Y_i - Y_i^l\right\|_2^2\right), \tag{31}$$

where $Y_i$ and $Y_i^l$ are, respectively, the $i^{\text{th}}$ value of the input data and predicted output along the loading paths that were taken in the experiments.

The degree of violation of the thermodynamic constraints is defined based on the regions where violations occurred in Figures 26-29e and 31e and is given by:

$$\text{Violation ratio} = \frac{A_V}{A_M}, \tag{32}$$

where $A_V$ is the area where violations occurred and $A_M$ is the total area of region that was modeled. The results are shown in Figure 32 for all the cases that have been considered here, from the basic neural network approach to the one that introduced Bayesian optimization.

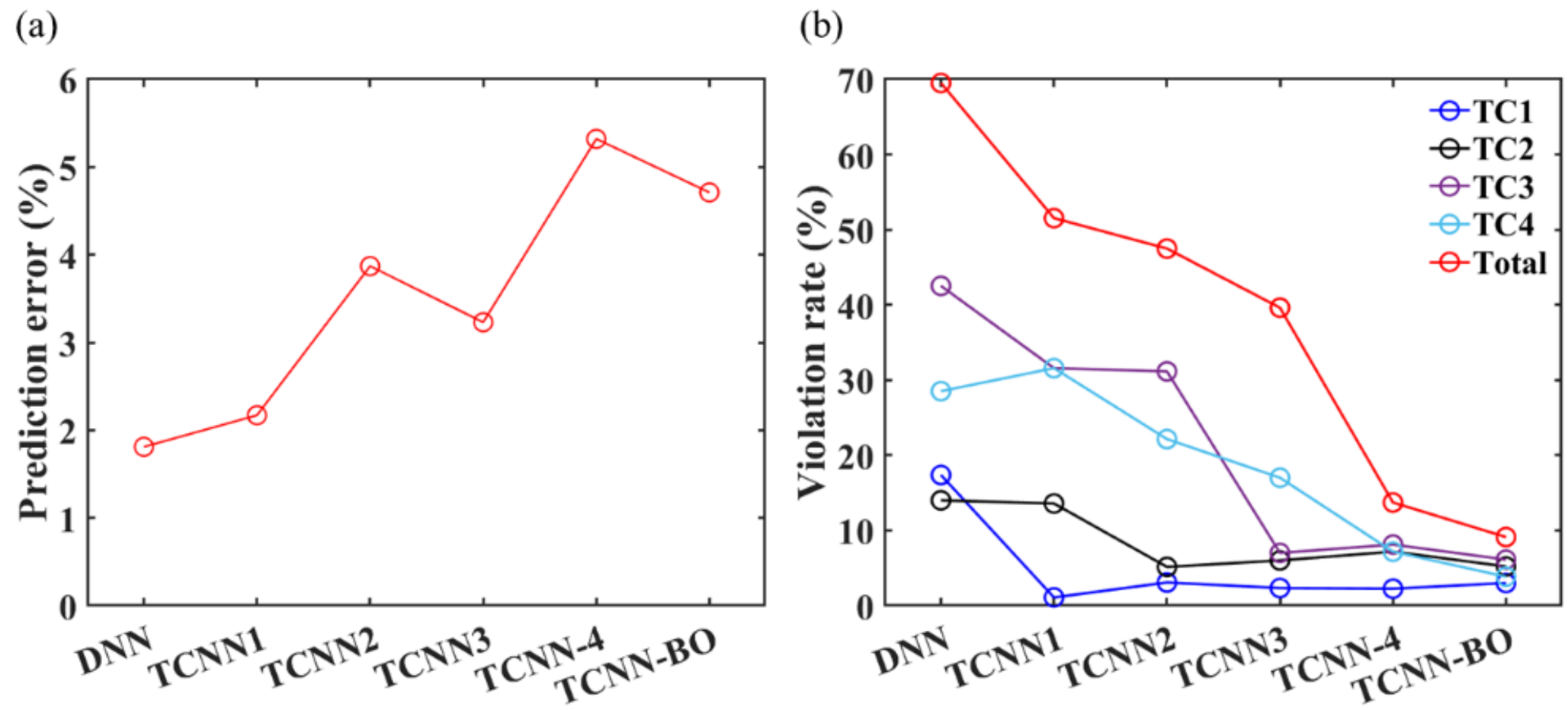

Figure 32. Comparisons of (a) the prediction errors and (b) violation rates across all six implementations of the neural network approaches that were considered here.

Interestingly the deep neural network, operating without any constraints, provided the smallest prediction error (Figure 32a). Furthermore, the neural network produced larger prediction errors as more TC conditions are introduced. Nonetheless, all prediction errors were less than 6 %, indicating that the input data was sufficiently well-learned by all five approaches.

By contrast (Figure 32b), the deep neural network operating by itself, without any constraints, had the highest violation ratio of all the approaches that were considered here. The violation rate across all four constraints was almost 70%, with most of that coming from the third and fourth constraints (TC3, TC4). As expected, this situation is significantly improved when the thermodynamic constraints were introduced as terms in the loss function. The first set of weight factors, which were directed at the first thermodynamic constraint, resulted in a significant reduction in its violation ratio, which had to be relaxed as the other constraints were introduced. This was due to the noise and systematic errors in the experiments, which induced additional conflict with the thermodynamic constraints. This conflict was mediated by introducing Bayesian optimization, which was able to strike a balance between capturing the main trend of the experimental data while bypassing the conflicts that are introduced by uncertainties in the experiments.

The errors presented so far are basically global errors. The local error reflects the difference between each data point on the traction surface and the corresponding point that was used to train the neural network. It is calculated separately for each of the constraints (TC1, TC2, TC3 and TC4) and then normalized with respect to the corresponding maximum values. The local error data is then organized into 10 bins from [0, 0.1] to [0.9, 1.0] and presented as a histogram (Figure 33) where the ordinate reflects the number of occurrences within each bin. The exercise was conducted for the neural network with and without thermodynamic constraints, culminating in the results from applying the Bayesian optimization to the thermodynamic constraints. Basically, the distribution becomes sharper as the constraints are systematically applied from Figure 33 a-e. The Bayesian selection of weight factors for the constraints provided the sharpest distribution of errors within 5%.

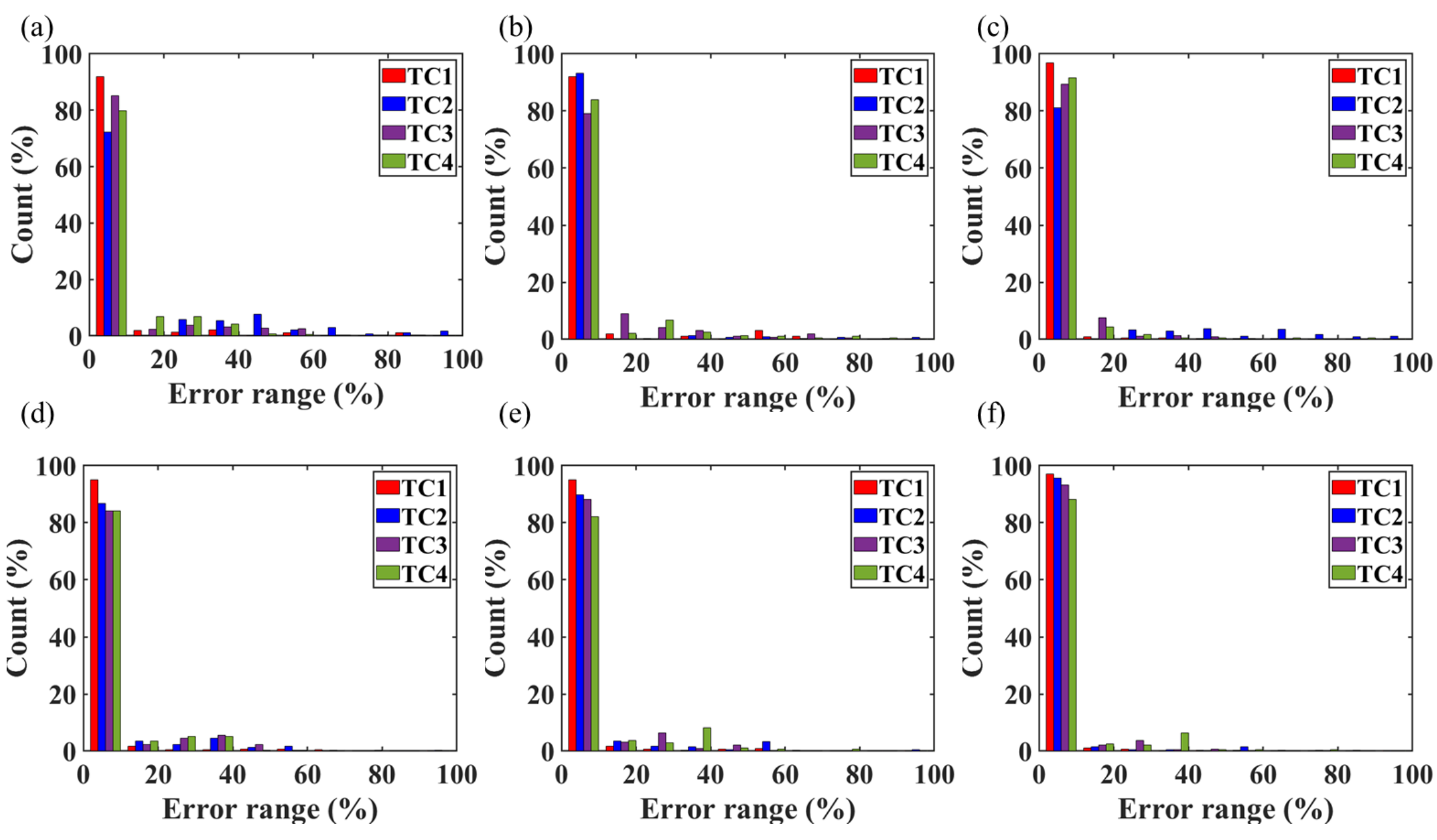

Figure 33. The distribution of the local error for modeling results with (a) no thermodynamic constraints, (b) MSE ($\lambda_0 = 0.9$) and TC1 constraint ($\lambda_1 = 0.1$), (c) MSE ($\lambda_0 = 0.8$) and TC1 and TC2 constraints ($\lambda_1 = 0.1, \lambda_2 = 0.1$), (d) MSE ($\lambda_0 = 0.78$) and TC1, TC2 and TC3 constraints ($\lambda_1 = 0.1, \lambda_2 = 0.1, \lambda_3 = 0.02$), (e) MSE ($\lambda_0 = 0.68$) and TC1, TC2 and TC3 constraints ($\lambda_1 = 0.1, \lambda_2 = 0.1, \lambda_3 = 0.02, \lambda_4 = 0.1$), (f) Bayesian optimization.

## 6. Conclusions

In this paper, we proposed a thermodynamically consistent neural network approach that provides interfacial traction-separation relations in the form of surfaces in traction *vs*. separation amplitude and phase angle space using experimental data that had been obtained along discrete loading paths. The proposed approach not only provides robust surfaces for mixed-mode traction-separation data but also conforms to thermodynamic principles that are directed at the development of damage once the strength of the interface has been exceeded. This is achieved by fusing the neural network model with thermodynamic consistency conditions as constraints. An additional constraint that reflects the variation of toughness with mode-mix was also considered. Furthermore, by integrating Bayesian optimization for the weight factors that reflect the priorities assigned to each constraint, the proposed approach was able to accommodate observed conflicts between the experimental data and thermodynamic consistency. While the prediction errors increased as constraints were systematically applied, they were still less than 5% once Bayesian optimization was applied to the selection of weight factors. In addition, the optimal weight factors also provided the lowest violation rate across all four constraints.

Moreover, a series of different training data sets were considered as subsets of the entire data set that was available, while the remaining data was used for validation purposes. Using training data along nine of the ten loading paths to predict the traction-separation relations along the tenth path was not very successful. Instead, using randomly selected sets of training data that consisted of 90% of the entire data set was generally more effective. Some of those training data sets were able to generate traction surfaces with errors that were within a few percent. These tended to be data sets that contained data from the damaging portion of the measured traction-separation relations.

It remains to be seen how effective cohesive zone modeling that embodies the generated surfaces are in predicting the strength and durability of multilayered components that are subject to interfacial failure under mixed-mode fracture conditions.

**Acknowledgements**

The authors would like to acknowledge the funding provided by the National Science Foundation under contracts #1930881 (Congjie Wei and Chenglin Wu) and EEC-1160494 (Kenneth M. Liechti). The latter is under the auspices of the National Science Foundation Nanosystems Engineering Research Center on Nanomanufacturing Systems for Mobile Computing and Mobile Energy Technologies (NASCENT). Jiaxin Zhang would like to acknowledge the funding supported by the U.S. Department of Energy, Office of Science, Office of Advanced Scientific Computing Research, Applied Mathematics program; and by the Artificial Intelligence Initiative at the Oak Ridge National Laboratory (ORNL). ORNL is operated by UT-Battelle, LLC., for the U.S. Department of Energy under Contract DEAC05-00OR22725.